\documentclass[]{spie}  

\usepackage{amsmath,amsfonts,amssymb,upgreek}
\usepackage{graphicx}
\usepackage{graphbox}
\usepackage{booktabs}
\usepackage{subcaption}
\usepackage[colorlinks=true, allcolors=blue]{hyperref}
\usepackage{fancyhdr}

\title{Conceptual Design of the Modular Detector and Readout System for the CMB-S4 survey experiment}

\author[a]{D.~R.~Barron} 
\author[b, c]{Z.~Ahmed} 
\author[d]{J.~Aguilar} 
\author[e, f, g]{A.~J.~Anderson} 
\author[h]{C.~F.~Baker} 
\author[i, j]{P.~S.~Barry} 
\author[k]{J.~A.~Beall} 
\author[i, f, g]{A.~N.~Bender} 
\author[e, f, g]{B.~A.~Benson} 
\author[d]{R.~W.~Besuner} 
\author[i]{T.~W.~Cecil} 
\author[i, f, g]{C.~L.~Chang} 
\author[l, m]{S.~C.~Chapman} 
\author[n]{G.~E.~Chesmore} 
\author[e]{G.~Derylo} 
\author[k]{W.~B.~Doriese} 
\author[k]{S.~M.~Duff} 
\author[d]{T.~Elleflot} 
\author[h]{J.~P.~Filippini} 
\author[e]{B.~Flaugher} 
\author[o]{J.~G.~Gomez} 
\author[p]{P.~K.~Grimes} 
\author[i]{R.~Gualtieri} 
\author[o]{I.~Gullett} 
\author[c]{G.~Haller} 
\author[b, c]{S.~W.~Henderson} 
\author[m]{D.~Henke}
\author[c]{R.~Herbst}
\author[q]{A.~I.~Huber} 
\author[k]{J.~Hubmayr} 
\author[e]{M.~Jonas} 
\author[d]{J.~Joseph} 
\author[r]{C.~L.~King} 
\author[p]{J.~M.~Kovac} 
\author[e]{D.~Kubik} 
\author[i]{M.~Lisovenko}
\author[d, f, n, g, s]{J.~J.~McMahon} 
\author[t]{L.~Moncelsi} 
\author[r, u]{J.~M.~Nagy} 
\author[h]{B.~Osherson} 
\author[c]{B.~Reese}
\author[o]{J.~E.~Ruhl} 
\author[c]{L.~Sapozhnikov}
\author[t]{A.~Schillaci} 
\author[e]{S.~M.~Simon} 
\author[d]{A.~Suzuki} 
\author[i]{G.~Wang} 
\author[v]{B.~Westbrook}
\author[i]{V.~Yefremenko}
\author[i]{J.~Zhang} 

\affil[a]{Department of Physics and Astronomy, University of New Mexico, Albuquerque, NM, USA}
\affil[b]{Kavli Institute for Particle Astrophysics and Cosmology, Menlo Park, CA 94025}
\affil[c]{SLAC National Accelerator Laboratory, Menlo Park, CA 94025}
\affil[d]{Lawrence Berkeley National Laboratory, 1 Cyclotron Rd, Berkeley, CA 94720}
\affil[e]{Fermi National Accelerator Laboratory, Batavia, IL 60510, USA}
\affil[f]{Department of Astronomy and Astrophysics, University of Chicago, 5640 S. Ellis Ave., Chicago, IL 60637, USA}
\affil[g]{Kavli Institute for Cosmological Physics, University of Chicago, 5640 S. Ellis Ave., Chicago, IL 60637, USA}
\affil[h]{Department of Physics, University of Illinois at Urbana-Champaign, Urbana, IL 61801}
\affil[i]{High Energy Physics Division, Argonne National Laboratory, 9700 S. Cass Av., Lemont, IL 60439}
\affil[j]{School of Physics and Astronomy, Cardiff University, The Parade, Cardiff, CF24 3AA, UK}
\affil[k]{National Institute of Standards and Technology, 325 Broadway, Boulder, CO 80305 USA}
\affil[l]{Dalhousie University, 6310 Coburg Rd., Halifax, NS  B3H 4R2 Canada}
\affil[m]{National Research Council, 5071 W Saanich Rd , Victoria, BC, Canada}
\affil[n]{Department of Physics, University of Chicago, Chicago, IL 60637, USA}
\affil[o]{Physics Department, Case Western Reserve University, 10900 Euclid Ave, Cleveland, OH, 44106, USA}
\affil[p]{Center for Astrophysics | Harvard \& Smithsonian, 60 Garden St, Cambridge, MA 02138}
\affil[q]{University of Victoria, Dept. of Physics and Astronomy, Victoria BC V8W 2Y2, Canada}
\affil[r]{Department of Physics,  Washington University in St. Louis, 1 Brookings Drive, St. Louis, MO 63130, USA}
\affil[s]{Enrico Fermi Institute, University of Chicago, Chicago, IL 60637, USA}
\affil[t]{Department of Physics, California Institute of Technology, Pasadena, California 91125, USA}
\affil[u]{McDonnell Center for the Space Sciences, Washington University in St. Louis, 1 Brookings Drive, St. Louis, MO 63130, USA}
\affil[v]{University of California, Berkeley, 151 Physics North, Berkeley, CA 94720}

\authorinfo{Further author information: (Send correspondence to D.~R.~Barron and Z.~Ahmed)\\D.~R.~Barron: E-mail: dbarron2@unm.edu\\  Z.~Ahmed: E-mail: zeesh@slac.stanford.edu}

\fancypagestyle{FirstPage}{

\lfoot{Copyright 2022 Society of PhotoOptical Instrumentation Engineers. One print or electronic copy may be made
for personal use only. Systematic reproduction and distribution, duplication of any material in this paper for a
fee or for commercial purposes, or modification of the content of the paper are prohibited.}

}
 
\begin{document} 
\maketitle

\begin{abstract}
We present the conceptual design of the modular detector and readout system for the Cosmic Microwave Background -- Stage 4 (CMB-S4) ground-based survey experiment. CMB-S4 will map the cosmic microwave background (CMB) and the millimeter-wave sky to unprecedented sensitivity, using 500,000 superconducting detectors observing from Chile and Antarctica to map over 60\% of the sky. The fundamental building block of the detector and readout system is a detector module package operated at 100\,mK, which is connected to a readout and amplification chain that carries signals out to room temperature. It uses arrays of feedhorn-coupled orthomode transducers (OMT) that collect optical power from the sky onto dc-voltage-biased transition-edge sensor (TES) bolometers.  The resulting current signal in the TESs is then amplified by a two-stage cryogenic Superconducting Quantum Interference Device (SQUID) system with a time-division multiplexer to reduce wire count, and matching room-temperature electronics to condition and transmit signals to the data acquisition system. Sensitivity and systematics requirements are being developed for the detector and readout system over wide range of observing bands (20--300\,GHz) and optical powers to accomplish CMB-S4’s science goals. While the design incorporates the successes of previous generations of CMB instruments, CMB-S4 requires an order of magnitude more detectors than any prior experiment. This requires fabrication of complex superconducting circuits on over 10\,m$^2$ of silicon, as well as significant amounts of precision wiring, assembly and cryogenic testing.  
\end{abstract}




\section{Introduction}
\label{sec:intro}  
\thispagestyle{FirstPage}

CMB-S4 is an upcoming survey experiment that will map the cosmic microwave background (CMB) and millimeter-wave sky to unprecedented sensitivity and precision~\cite{cmbs4_sciencebook}. It will enable a wide range of science in cosmology and astrophysics, and carries the potential to transform our understanding of the universe. First conceived by the community during the 2013 Snowmass physics planning activity as the ultimate ground-based CMB survey, CMB-S4 builds on several prior generations of CMB experiments. The motivation and potential impact of CMB-S4 was also described in the Astro2020 decadal survey, which ranked it as a top priority for the next decade\cite{astro2020}.
The science case for CMB-S4 includes searching for primordial gravitational waves predicted by cosmic inflation;
searching for the effects of new light relic particles in the early universe; mapping the matter distribution of
the universe; and opening a new window on the millimeter-wave, time-variable astronomical sky.
These exciting science goals require an exceptionally deep survey to hunt for the faint signal from cosmic inflation, and a precise, high-resolution survey of the majority of the sky to measure as many spatial modes of the CMB as possible.
%
CMB-S4 will therefore conduct two surveys; one targeting 3\% of the sky, sensitive to both degree angular scales and arcminute angular scales to search for the signal from cosmic inflation, and a second targeting 60\% of the sky sensitive to arcminute angular scales to search for the effects 
of light relic particles, map the matter density, and do transient millimeter-wave astronomy. 
The survey requirements drive CMB-S4 to use 5-6\,m diameter telescopes, dubbed ``Large Aperture Telescopes" (LATs) to achieve arcminute angular resolution, and 0.5-m diameter telescopes (``Small Aperture Telescopes") (SATs) to measure the degree-scale signals with lower instrumental systematic errors.

CMB-S4's science goals require 500,000 photon-noise-limited detectors in these telescopes, a significant increase compared to prior experiments. 
The measurements must also be made across a decade in observing frequencies in order to 
separate the CMB signal from Galactic foregrounds. 
%
CMB-S4 will use the same modular detector and readout electronics implementations in the focal planes of the LATs and SATs, differing only where necessary to achieve optimal performance and production efficiency.
The large detector count necessitates robust and scalable methods for fabrication and packaging the detectors and cryogenic readout components; this has influenced the conceptual design for the system. Components must be tested and validated to meet stringent performance requirements, and re-use and re-working of components is expected to be necessary to yield integrated modules meeting deployment criteria.

In this proceedings, we describe the conceptual design for the modular detector and readout systems that will be used in CMB-S4. 
In Section \ref{sec:requirements}, we discuss the development of technical requirements that flow down from the science goals of the experiment, and the key requirements that drive the design for the detector, readout, and module sub-systems. 
In Section \ref{sec:concept}, we describe the high-level detector and readout concept including technologies utilized and technical choices made in defining the modular detector and readout system. 
In Section \ref{sec:modularimplementationfors4}, we describe the specific implementation of these technologies for CMB-S4 including the details of the modular design. 
In Section \ref{sec:test_stands}, we discuss the development and design validation plan, and in Section \ref{sec:production}, we describe future work on scaling production of these components.

\section{Requirements for CMB-S4 Detector \& Readout System}\label{sec:requirements}

The requirements and performance targets of the CMB-S4 detector and readout system are derived from an iterative process of systems engineering that flows the experiment's science goals to measurement and technical requirements, and then down to requirements on subsystems and individual assemblies and components, as described in Besuner et al. in these proceedings. We report the current state of these requirements, which we continue to mature.  An important consideration for technical implementation decisions is technical readiness of potential designs.  Minimizing technical risk is prioritized, leading to frequent choosing of proven approaches or low-risk variations on them. Details of technical choices are discussed in \ref{sec:concept} and the implementation and resulting target requirements are discussed in Section \ref{sec:modularimplementationfors4}.
The overall project system requirements (Level 1 Technical Requirements) that impact the detector and readout system are driven largely by the instantaneous sensitivity needed to measure CMB temperature and polarization at various observing frequencies, at sufficient spatial resolution while scanning from CMB-S4 telescopes, to achieve the necessary map depths (Level 1 Measurement Requirements) within $\sim$7 years of science operations, while keeping systematic errors subdominant to statistical uncertainty. This translates roughly to requirements on observing-band-wise counts of photon-noise-limited, polarization-sensitive detectors to integrate sky signal under historical optical loading conditions and assumed observing efficiency at the sites. 

Next-level-down derived requirements (Level 2 Subsystem requirements) are assigned to the three Level 2 subsystems of the project, Detectors, Readout and Modules, encompassing the detector and readout system. Target values are prescribed for detector counts by observing band, band center and edge frequencies, pixel operability or channel yield, per-detector noise-equivalent power and its acceptable readout contribution, supported dynamic range of input optical power (saturation power), detector time constant and readout bandwidth. Based on technical choices described in Section \ref{sec:concept}, such as the choice of transition-edge-sensor (TES) bolometers, orthomode transducers (OMT) and time-division multiplexing (TDM) as well as interfaces between the detectors and readout, further technical targets are prescribed such as TES normal resistance, superconducting transition temperature, orthomode transducer orientation, inter-channel crosstalk, electro-thermal feedback loopgain, readout sampling and multiplexing rates, etc. There is also a requirement for an additional in-series TES of higher saturation power for every polarization sensor to enable optical calibration with external sources.
The current target values for some of the system's key requirements are enumerated in Table~\ref{tab:requirements}.  These targets are the result of the project's iterative approach, and continue to be optimized to work with the other subsystems of the experiment to meet the overall project requirements. Some requirements described above are still being developed. They provides a basis for system prototyping, which is underway and described in Section \ref{sec:prototyping}.

\begin{table}[]
\scriptsize
\renewcommand{\arraystretch}{1.5}
\begin{tabular}{l|cccc|cccc}
                                              & \multicolumn{4}{c}{\textbf{Large-aperture telescope (LAT)}}                                                                                                     & \multicolumn{4}{|c}{\textbf{Small-aperture telescope (SAT)}}                                                                                                            \\
\textbf{Detector Wafer Type}                          & \textit{ULF}             & \textit{LF}                       & \textit{MF}                       & \textit{HF}                       & \textit{LF}                       & \textit{MF1}                     & \textit{MF2}                     & \textit{HF}                       \\
                                              \hline
\textbf{Band center(s) {[}GHz{]}}             & 20 & 26 / 39 & 93 / 149 & 227 / 286 & 26 / 39 & 85 / 145 & 95 / 155 & 227 / 286 \\

\textbf{Fractional bandwidth}                 & 0.25                     & 0.33 / 0.45                       & 0.32 / 0.28                       & 0.26 / 0.21                       & 0.33 / 0.45                       & 0.24 / 0.22                      & 0.24 / 0.22                      & 0.26 / 0.21                       \\
\textbf{Saturation power {[}pW{]}}                        & 0.40                     & 0.75 / 4.2                        & 4.6 / 13                          & 32 / 42                           & 1.4 / 6.4                         & 7.9 / 14                         & 7.9 / 14                         & 33 / 41                           \\

\textbf{Dark NEP {[}aW/$\boldsymbol{\sqrt{Hz}}${]}}                        & 5                     & 6 / 22                        & 22 / 42                          & 81 / 103                           & 13 / 34                         & 42 / 57                         & 41 / 59                         & 109 / 133                           \\

\hline
    \textbf{Optical efficiency}                                          & \multicolumn{8}{c}{$65\%$ average in-band efficiency (for detector module only)}  \\  

    \textbf{Operable channels}                                          & \multicolumn{8}{c}{$\geq 85\%$ per wafer, installed, including losses from detectors, wirebonds, readout}    \\
    
    \textbf{Transition temperature}                                          & \multicolumn{8}{c}{160 mK (science TES)}    \\
    
\hline

\end{tabular}
\vspace{1mm}

\caption{Current target parameters for CMB-S4's detector and readout system by observing band groups or wafer type. All  except \textit{ULF} have dichroic pixels. The saturation power shown is for the science TES. Dark noise-equivalent power (NEP) includes expected noise contributions from the detector and readout system.
}
\label{tab:requirements}
\end{table}

\section{Detector \& Readout System Concept}\label{sec:concept}


The CMB-S4 detector and readout system concept employs arrays of TES bolometers, coupled to the sky via feedhorns and read out using time-division multiplexing (TDM) of Superconducting Quantum Interference Devices (SQUIDs).
These design choices are the product of an analysis of the available technologies, weighing their demonstrated performance, estimated cost and risk, and comparing to the Level 1 Technical Requirements from the science flowdown.
TES bolometers are a well-established technology across CMB-S4's entire frequency range, with demonstrated noise levels and fabrication yields that meet the instrument requirements~\cite{westbrook_2018,DuffYield,sobrin_spt3g_2022,Moncelsi2020,henderson_advact_2016}. 
Feedhorns coupled to planar ortho-mode transducers (OMTs) have been demonstrated to provide excellent beam quality and polarization efficiency, with recent advances in direct machining and electromagnetic optimization enabling high-quality horns to be manufactured economically at scale.
Each CMB-S4 feedhorn will deliver power to four TESs: two orthogonal polarization modes for each of two frequency bands, defined by on-chip lumped-element filters; this dichroic architecture allows for high sensor density while maintaining high end-to-end optical efficiency in all observing bands.
TDM readout has a long heritage in CMB instrumentation, with sufficiently demonstrated high yield and low noise (both white and $1/f$), and with recent developments for X-ray micro-calorimetry enabling higher multiplexing factors and lower noise levels\cite{henderson_readout_2016,Moncelsi2020}.
In this section, we describe each of these major design components in more detail.

\subsection{Optical Coupling}
\begin{figure}[h!]
\centering
\includegraphics[width=0.6\textwidth]{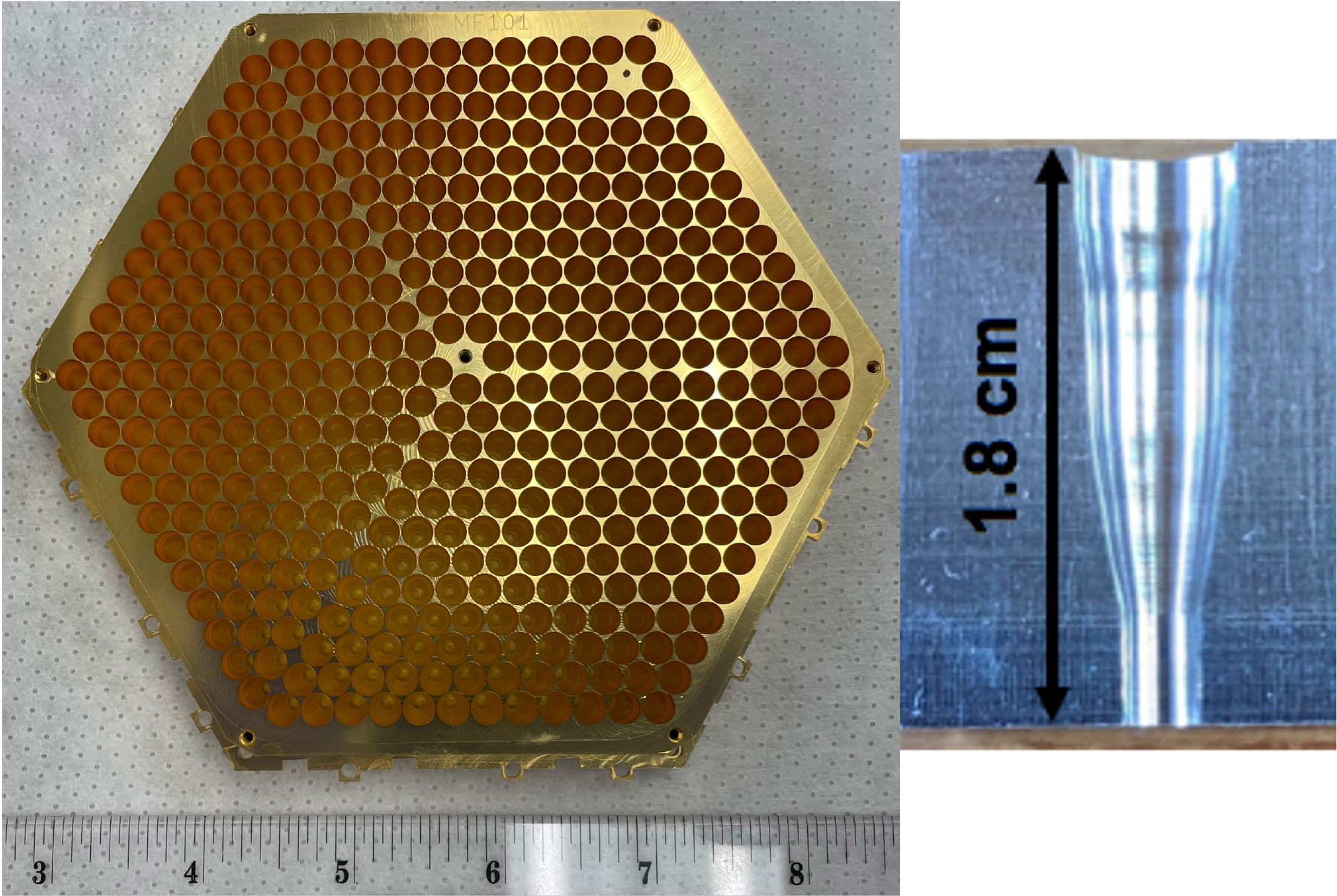}
\caption{The left shows a gold-plated Al feedhorn array for 430 optical pixels of 90/150\,GHz from Simons Observatory (SO), while the right shows a cross-sectional profile of a single direct-machined feedhorn with a spline profile. CMB-S4 will use this feedhorn profile and arraying scheme.}
\label{fig:horn}
\end{figure}

The detector antennae are feedhorn-coupled planar orthomode transducers (OMTs). The feedhorns fully define the detector beams. We optimize the feedhorn performance to the specific optical requirements of the telescope and frequency band using Markov Chain Monte Carlo methods~\cite{spline_horns}. Requirements typically include edge taper, ellipticity, and/or spillover. Previous experiments have typically used stacks of $\sim$40 through-etched Si wafers to build up the feedhorn profiles in an array~\cite{ACTPol_Instrument}. However, new methods using direct-machining into Al with custom tooling can reduce the time and cost of production by a factor of $\sim$20~\cite{horn_fab}. The feedhorns are designed to have a monotonically increasing profile shape to enable direct machining~\cite{Zeng2010}. A direct-machined feedhorn array and single feedhorn cross-section are shown in Figure~\ref{fig:horn}.

\begin{figure}[h!]
\centering
\includegraphics[width=0.98\textwidth]{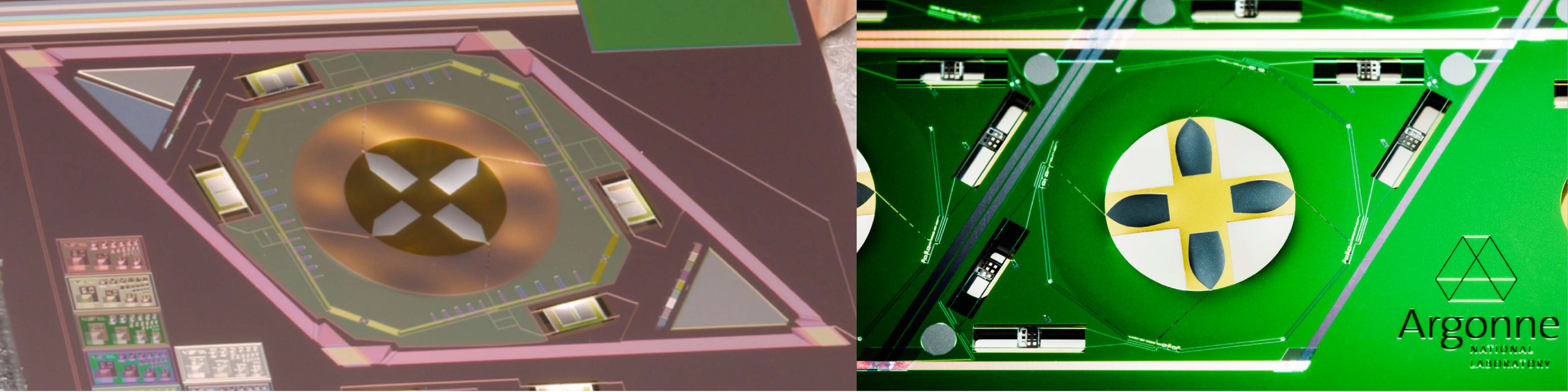}
\caption{The OMT sits in the center of each pixel and is comprised of four fins. {\it Left:} A pixel fabricated by the National Institue of Standards and Technology, Boulder (NIST) with an OMT design from previous CMB experiments {\it Right:} A CMB-S4 prototype pixel fabricated by Argonne National Laboratory (ANL) with the new ``wine glass'' probe OMT design.}
\label{fig:OMT}
\end{figure}

The OMT consists of four Nb probes on a low-stress, SiN membrane that split the polarization into two orthogonal directions.  Previous experiments have typically used probes with linear features (left panel of Figure~\ref{fig:OMT})~\cite{OMT_TRUCE}, but CMB-S4 will use a new probe design with a ``wine glass" shape (right panel of Figure~\ref{fig:OMT}) that has a more uniform response in frequency and a $\sim$2\% efficiency gain in the high band. This new design was made possible by improved computing power enabling more complex numerical optimizations. The OMT design was optimized for the LAT mid-frequency detector wafer (MF), and this OMT design is linearly scaled for other observing bands with a scaling factor optimized for the bands.
The radiation from the probes is passed into a superconducting co-planar waveguide (CPW). A stepped impedance transformer is employed to transition to the low impedance microstrip lines that make up the on-chip detector filters and circuitry.

\subsection{Transition of optical power to sensor}

After the stepped impedance transformer, the optical signal is transmitted through superconducting microstrip transmission line. The microstrip consists of a ground layer, which is typically niobium (Nb); a dielectric spacer, either a silicon oxide (SiO$_\textrm{x}$) or a silicon nitride (SiN$_\textrm{x}$); and a top conductor, usually the same material as the ground layer. The signal is routed to an in-line diplexer that partitions the signal into the two optical passbands. The diplexer filter uses either a lumped element or stub filter components. After the diplexer, the signal is coupled to the bolometer through one of two approaches. One approach terminates a pair of transmission lines (each connecting to a one of the opposing OMT fins) across a matched impedance load resistor. If the input line lengths are equal, then only in-band power from the TE$^\circ_{\textrm{11}}$ is dissipated in this configuration. The other approach feeds the pair of microstrip lines into a hybrid tee, which produces a sum and a difference output from the two input signals. The higher order modes from the sum output are terminated on the substrate and the TE$^\circ_{\textrm{11}}$ mode from the difference output is routed to the bolometer where the signal is dissipated through a sufficient length of lossy material. In places where microstrip transmission lines are required to cross each other, the design employs cross over structures either using vias or additional dielectric and conductor material.

\subsection{Transition-Edge Sensor bolometer}
The superconducting microstrip is terminated on   
Transition-Edge Sensor (TES)\cite{1995ApPhL..66.1998I} bolometers, which transduce changes in optical power to changes in current. TES bolometers have been adopted widely by recent CMB experiments for their scalable manufacturing\cite{1999ApPhL..74..868G,DuffYield,2018JLTP..193..703P}, well-understood noise properties\cite{2005cpd..book...63I} and 
ability to achieve ``background-limited'' operation where detector noise is dominated by 
photon shot noise. \cite{westbrook_2018,DuffYield,sobrin_spt3g_2022,Moncelsi2020,henderson_advact_2016} 
The detectors are voltage biased, with strong 
electro-thermal feedback 
resulting in a highly linear response with improved response time
\cite{1996ApPhL..69.1801L}.
As a low impedance sensor, the TES is compatible with several multiplexing readout technologies. 

\begin{figure}
    \centering
    \includegraphics[width=2in,align=c,vshift=0.1in]{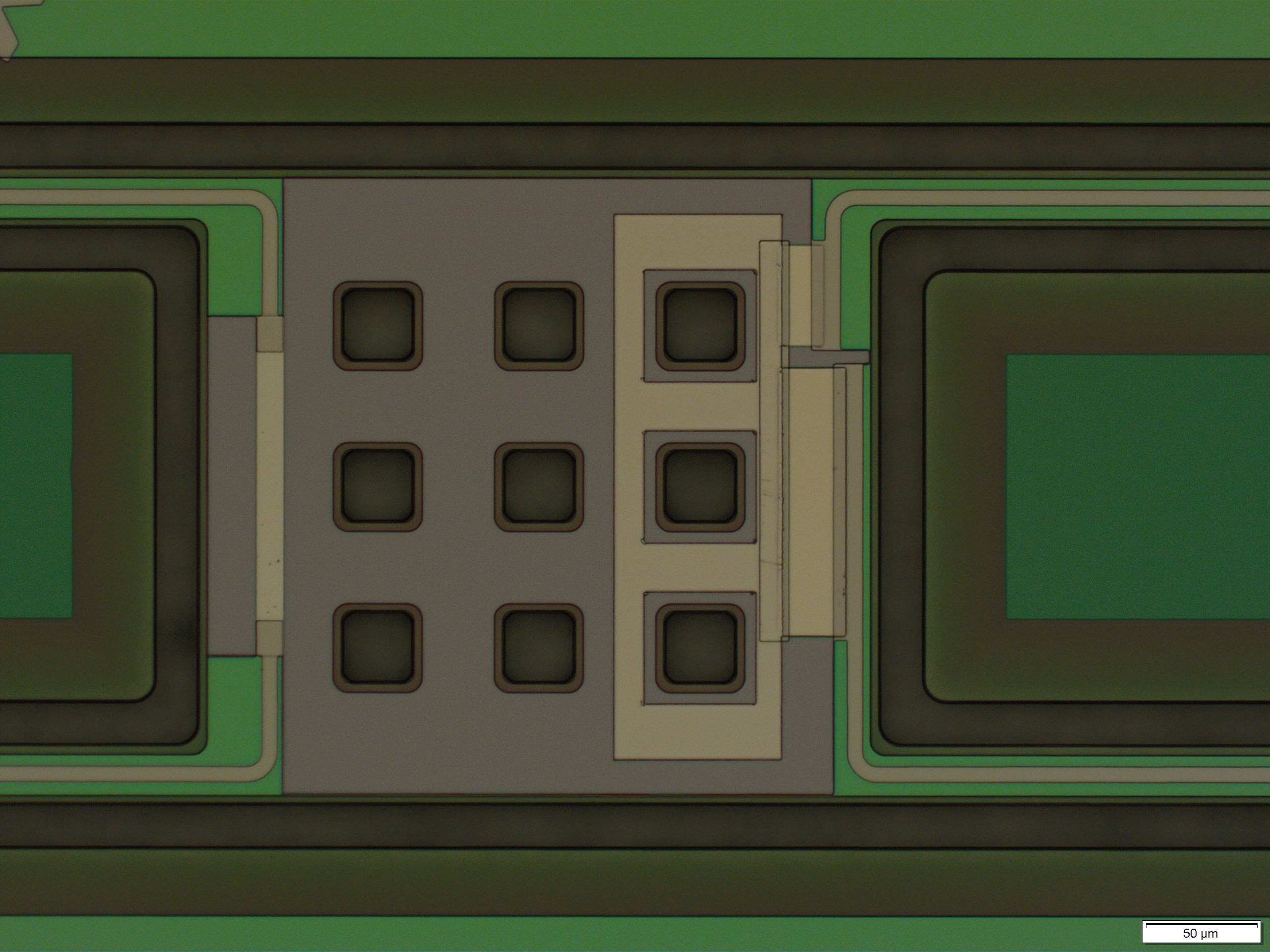}%
    \hspace{0.1in}
    \includegraphics[scale=1,align=c]{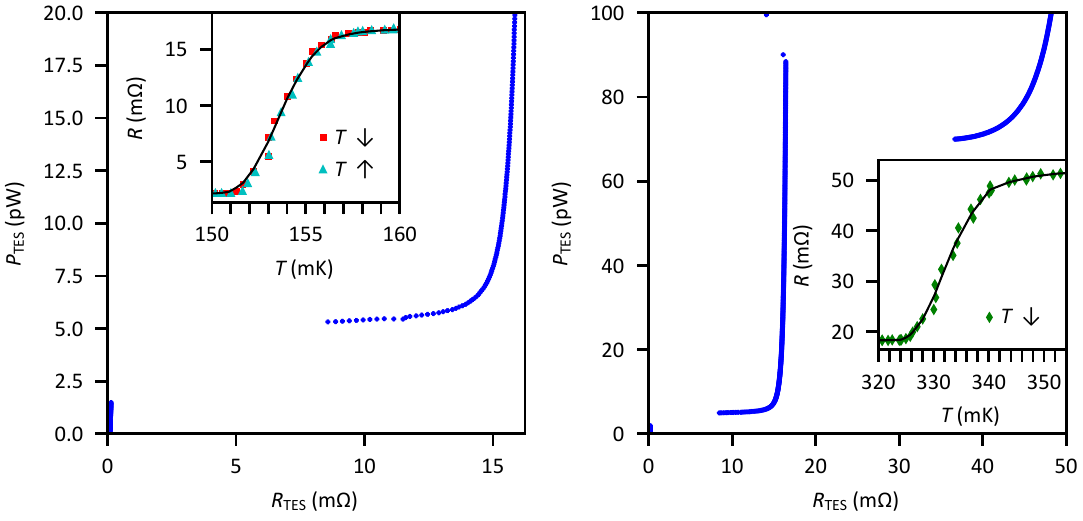}
    \caption{{\it Left:} Photo of an prototype CMB-S4 bolometer island. The optical signal is differentially terminated across a matched load resistor on the left side of the island. The right side of the island has the calibration TES (smaller TES, upper right) and science TES (larger TES, center right). The island is suspended from the silicon substrate and thermal transport along the legs determine the bolometer primary thermal conductance. Additional heat capacity is added to tune the detector time constant. {\it Center:} Plot of measured bias power versus detector resistance for the science TES for a prototype CMB-S4 detector. The TES resistance, $R_\textrm{TES}$, transition temperature, $T_c$, (inset), and saturation power, $P_\textrm{sat}$ are approaching development targets. {\it Right:} Plot of measured bias power versus detector resistance for the calibration TES (upper right branch; left branch corresponds to the science TES in center panel). The TES resistance and transition temperature are near development targets and the large saturation power is sufficient for the calibration measurements.}
    \label{fig:TES}
\end{figure}

CMB TES bolometers are fabricated using thin-film microfabrication techniques used in both superconducting microelectronics and MEMs applications, which enables scaling to large array production. The key parameters for the TES bolometer designs are the superconducting transition temperature of the sensor ($T_c$), sensor impedance ($R_{op}$), the bolometer thermal conductance ($G$), or equivalently saturation power, $P_\textrm{sat}$, 
and the time constant ($\tau$) of the bolometers. 
Figure \ref{fig:TES} shows a picture of a CMB-S4 prototype TES bolometer.
Each bolometer will have two TESs fabricated on the same released area of the bolometer and connected in series\cite{2010SPIE.7741E..0HO}. The two TES have different superconducting transition temperatures, with one optimized for science observations and the other designed for higher-power calibration sources. 
For the science TES, $T_c$ is chosen as $\sim$160\,mK and $G$ is chosen band-wise to minimize the phonon-carrier noise of TES bolometers for an operating temperature of 100\,mK, while providing sufficient dynamic range or operating margin under the predicted optical load for that band without saturation. 
$P_\textrm{sat}$ is the total power dissipated when the bolometer is voltage biased at the operating point. When observing the sky, this power consists of both 
incoming optical power and electrical bias power. 
For CMB-S4, $P_\textrm{sat}$ is designed to be roughly a factor of three greater than the expected optical power providing margin 
for changing optical loading due to changes in weather and the telescope's observing elevation. 
The normal resistance ($R_\textrm{norm}$) of the science TES is currently targeted at $14\pm2$ milli-Ohm. The $T_c$ for the calibration TES is higher than the science TES, and chosen so that the detector can observe a 450\,K calibration source without saturating. The calibration $R_\textrm{norm}$ is chosen so that it can be reliably voltage biased when including the normal series resistance of the science TES.
The upper bound of the bolometer $\tau$ is chosen such that sensors go through multiple $\tau$ periods as the detector beam scans a physical scale of interest across the sky. The lower bound is set by readout bandwidth and electro-thermal-feedback stability. 
CMB-S4's science TES bolometer $\tau$'s will have a lower bound of 1\,ms and an upper bound of $\sim 3 - 37$\,ms depending on the observing band.

The science TES will consist of a sputtered film of aluminum manganese (AlMn) alloy\cite{2004ApPhL..85.2137D,2011ITAS...21..196S}.
The Mn concentration and the temperature of a controlled bake of the AlMn films are used to set the  $T_c$\cite{2016JLTP..184...66L}. 
Both the lateral dimension and thickness of the TES are chosen to achieve the required $R_\textrm{norm}$. 
A film of gold, gold/palladium, or palladium, grown either by sputtering or electron beam evaporation, will be used as a thermal anchor where the volume of this heat capacity is chosen to provide the required $\tau$.
Niobium film is used for the superconducting leads to connect to the TESs and the RF transmission lines.
All of the TES structures are fabricated 
on a low stress SiN$_\textrm{x}$ membrane, which is then patterned and released to form narrow bridges suspending the TES island. The release is carried out either via backside through wafer etch with a deep reactive ion etching (DRIE) or a xenon fluoride process.
The bolometer $G$ is determined either by the electrons in a strip of normal metal along one of the bridges, or by the phonon transport along the bridges. In the case of the latter, the ratio of cross-section area versus length of the suspended bridges is adjusted to tune thermal conductance. 


\subsection{Transition-edge sensor biasing}

A direct-current (DC) voltage bias is applied to each TES bolometer to heat it into its superconducting transition.
CMB-S4 will apply these biases using a small ($\sim 400\,\upmu\Omega$) shunt resistance wired in parallel with each TES as shown in Figure~\ref{fig:mux_column_schematic_incl_tes_biasing}, driven by differential current sources in the warm electronics selected for low $1/f$-noise.
A single bias line can drive several dozen TES/shunt pairs wired in series, typically corresponding to a single column of the TDM readout.
The use of shared bias lines requires percent-level tolerances on uniformity among detectors in a wafer, such that all detectors on a given bias line can simultaneously achieve near-optimal performance. 
In parallel with its shunt resistor, each TES is also wired in series with its associated SQUID input coil and with a discrete series inductor.
The total inductance in this loop (discrete, input coil, and stray) is tuned to control the bandwidth of the TES circuit.
This ensures stable operation of the TES, and controls aliasing of TES and SQUID noise from the sampling rate of the TDM system.
This introduces important couplings between design parameters of the TESs and readout system, which must be accounted for in system design.

\subsection{\label{ssec:sqampchainandmux} SQUID amplifier chain and multiplexing}

\begin{figure}[h!]
\centering
\includegraphics[width=0.95\textwidth]{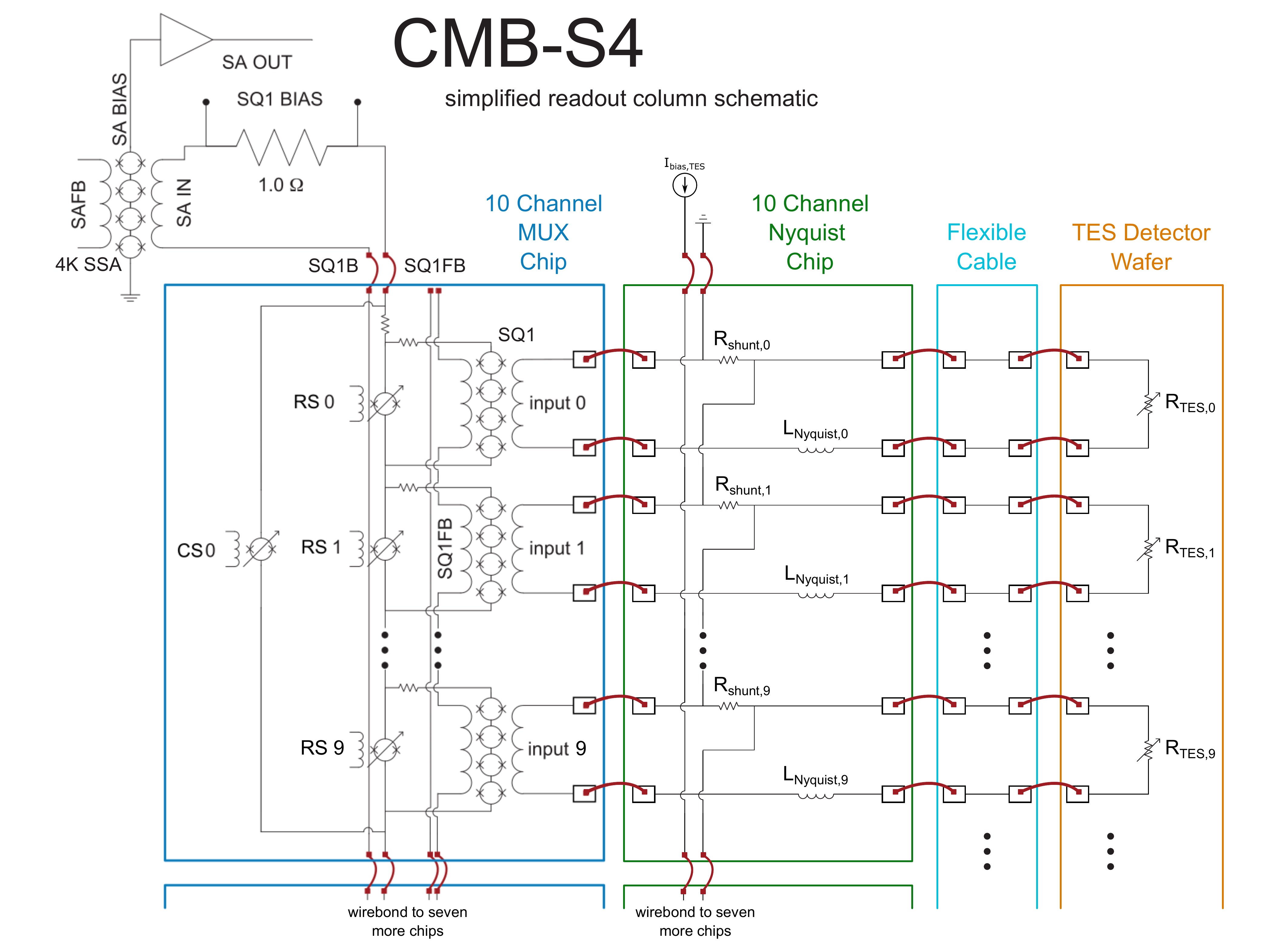}
\caption{System-level schematic of the TES biasing, SQUID amplification and multiplexing circuit for one TDM readout column for CMB-S4.  For simplicity, not all cables and interconnects are shown.  
Each TES, represented by $\mathrm{R_{TES}}$, in a CMB-S4 detector wafer is connected through a superconducting flexible circuit using pairs of superconducting aluminum wirebonds into one of the inputs of a 10-channel Nyquist chip which adds a wirebond selectable inductance in series with the TES $\mathrm{L_{Nyquist}}$ and a bias resistor shunting the TES $\mathrm{R_{shunt}}$.  Another pair of superconducting aluminum wirebonds puts the TES in series with the input of a dedicated first stage SQUID series array amplifier (SQ1) on a 10-channel multiplexing (MUX) chip.  One TES at a time is read out on a column by switching its SQ1 into the column readout circuit by activating its MUX chip's row select (RS) and chip select (CS) as described in the text.  The common column readout circuit for every SQ1 consists of a second stage SQUID series array amplifier at 4K, followed by a warm preamplifier at room temperature.  In time division multiplexing, every TES is read out periodically by cyclically activating its CS and RS switches and sampling the output of the column readout circuit.}
\label{fig:mux_column_schematic_incl_tes_biasing}
\end{figure}

CMB-S4 will use an implementation of time-division multiplexing (TDM) using DC SQUIDs to read out its TES bolometers. This technology has over a decade of heritage on fielded CMB receivers \cite{irwin_tdm_2002}. In TDM, TESs are read out in a 2D grid of rows and columns.  The current signal from each TES is first amplified by a dedicated first-stage SQUID (SQ1), which is shunted by a Josephson junction (JJ) switch.  Many SQ1s shunted by switches are chained in series and connected to the input of second stage SQUID Series Array (SSA) at $\sim$4 K.  A common SQ1 feedback coil is inductively coupled to every SQ1 in a column in series to allow operating any SQ1 in a closed flux-locked loop.  Flux coupled to any JJ switch through an inductively coupled control line can either make the switch  superconducting (off), shorting out its SQ1, or resistive (on), exposing its SQ1 to the column’s SSA input, depending on the flux applied.  Switch control lines are connected in series for one switch from each column to form a readout row, controlled by a single row-select (RS) line.  TES arrays are then read out by switching on one row at a time with all others off while operating the on row’s SQ1s in a closed flux-locked loop.  This TDM architecture was first deployed on BICEP3 in 2015\cite{bkxv_2022}, and subsequently on many other instruments.

To use TDM in CMB-S4 we aim to increase scalability, and reduce the cost and integration complexity by increasing the multiplexing factor, or number of TESs that can be read out per column.  In prior TDM readout implementations on CMB instruments, the multiplexing factor has been limited by the achieved readout bandwidth.  The readout bandwidth limits the row switching rate, which in turn results in a degradation of noise performance due to SQUID and TES noise aliasing as the multiplexing factor is increased.  The AdvACT experiment has demonstrated the highest multiplexing factor to date using the legacy TDM architecture\cite{henderson_readout_2016}, reading out 64 rows on 32 columns with a row switching rate of 2~$\upmu$s resulting in an increase in noise due to aliasing of between 5-10\%\cite{gallardo_aliasing_2020}.  To increase the multiplexing factor we will incorporate several low-risk technological improvements that have been developed by NIST Boulder for the readout of much faster X-ray TESs\cite{doriese_tdm_2016,durkin_bandwidth_2021,smith_tdm_2021}.  Planned improvements include adding a shunt resistor across the SSA to increase the system bandwidth\cite{zeng_shunt_2013}, a new faster SSA design, and a new SQ1 design with a higher input mutual inductance.  Taken together, preliminary studies indicate these improvements may enable substantially lower row switching rates and thereby multiplexing factors in excess of 120 rows per column, but we baseline 80 rows in the conceptual design.  

Additionally, we will incorporate new two-level switching SQUID multiplexing architectures\cite{dawson_two-level_2019} into CMB-S4's detector readout to significantly reduce the number of wires required to switch the rows. These architectures connect banks of RS switches in parallel with a single control switch shunting each bank to significantly reduce the number of required wires.  One switch from each bank is connected serially across all banks and columns to form a two-dimensional switching matrix.  To address an individual row in the multiplexer, warm electronics must activate both the desired RS switch, and the control switch for the bank the row is in, or the chip select (CS) switch.  
A schematic of the readout circuit for one column of the CMB-S4 multiplexing architecture is shown in Figure~\ref{fig:mux_column_schematic_incl_tes_biasing}.  Prototype chips implementing this new architecture are being fabricated now at NIST Boulder and will be tested soon in CMB-S4 detector and readout cryogenic test stands.

\section{Modular Implementation for CMB-S4}\label{sec:modularimplementationfors4}

\begin{figure}[h!]
\centering
\includegraphics[width=0.95\textwidth]{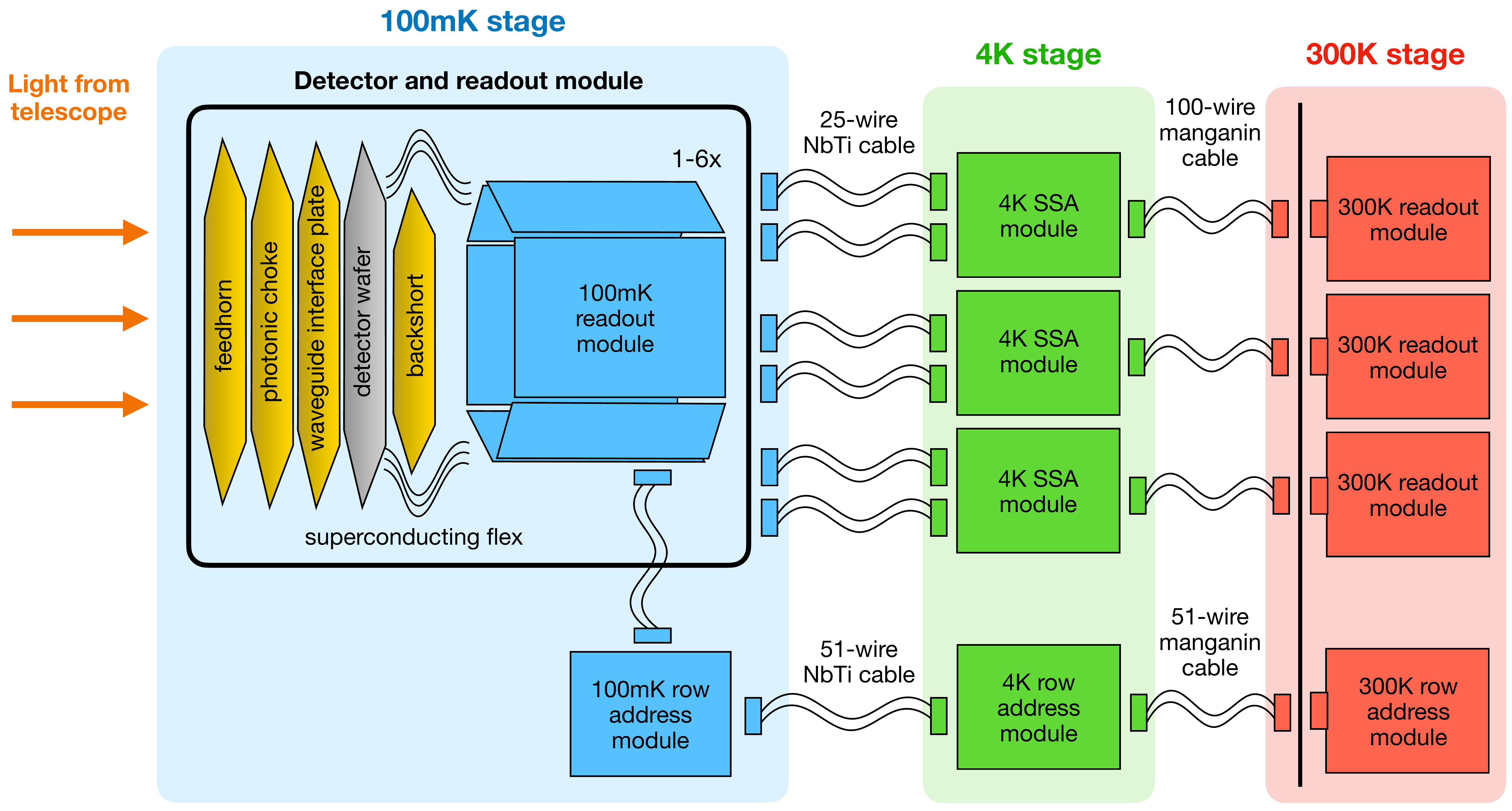}
\caption{System-level schematic of modular readout scheme showing components at each temperature stage and interconnects. Light from the telescope optics is incident on the optical coupling components and detectors, which are coupled to readout modules at 100\,mK, and comprise the integrated detector and readout module.
Cabling connects the 100\,mK readout to boards at 4\,K containing SQUID series arrays, which are connected to room-temperature readout electronics outside the cryostat.
Row addressing signals are carried by another series of cables and modules that pass from the room-temperature electronics to the 100\,mK readout modules.}
\label{fig:drm_schematic}
\end{figure}

The concrete implementation of detectors, readout electronics, and optical coupling in CMB-S4 emphasizes modularity as an overarching design guideline, which facilitates quality control during production, and enables reuse and reworking of components.
A schematic of the modular implementation concept is shown in figure \ref{fig:drm_schematic}. Silicon optical coupling wafers are assembled into a stack together with a feedhorn array and a single 150-mm silicon wafer of TES bolometers (``detector wafer''), to form the waveguide that efficiently couples light from the telescope into the TESs via OMTs and microstrip.
This optical stack is mounted to a frame that also contains a number of 100\,mK readout modules, consisting of the first-stage SQUID amplifiers, TES bias resistors, Nyquist-bandwidth-defining filters, and wiring on silicon and printed circuit boards.  Each 100\,mK readout module is interchangeable and consists of 4 readout columns of an 80-row TDM architecture with two-level addressing.
The 100\,mK readout modules connect to the detectors via a superconducting flexible circuit, as well as with NbTi cables to the SQUID series arrays at 4\,K which form the next stage of the amplification chain in the readout. The chain terminates in warm readout electronics modules mounted outside the camera cryostats, and connected to 4\,K readout elements with manganin cables.

The modular design balances the need for optimal sensitivity and a desire to use interchangeable components conducive to mass production and testing.
Detector parameters, wafer layouts, and feedhorn designs are optimized for each telescope type (LAT and SAT) and observing band to meet L2 subsystem requirements, L1 technical requirements, and up through the measurement requirements and the science goals. 
%
This results in eight unique detector and feedhorn designs, whose basic parameters are summarized in Table~\ref{tab:ubertable}.

\begin{table}[]
\begin{tabular}{l|cccc|cccc}
                                              & \multicolumn{4}{c}{\textbf{Large-aperture telescope (LAT)}}                                                                                                     & \multicolumn{4}{|c}{\textbf{Small-aperture telescope (SAT)}}                                                                                                            \\
\textbf{Detector wafer type}                      & \textit{ULF}             & \textit{LF}                       & \textit{MF}                       & \textit{HF}                       & \textit{LF}                       & \textit{MF1}                     & \textit{MF2}                     & \textit{HF}                       \\
                                              \hline

\textbf{Number of wafers}                     & 4                        & 17                                & 108                               & 41                                & 24                                & 72                               & 72                               & 48                                \\
\textbf{Active detectors / wafer}             & 54                       & 192                               & 1728                              & 1876                              & 48                                & 588                              & 676                              & 1876                              \\
\textbf{Total active detectors}               & 216                      & 4800                              & 279,936                            & 120,064                            & 1152                              & 36,288                            & 41,760                            & 68,736                             \\
\hline
\textbf{Readout modules / wafer} & 1                        & 1                                 & 6                                 & 6                                 & 1                                 & 3                                & 3                                & 6                                 \\
\textbf{Readout columns / wafer}              & 4                        & 4                                 & 24                                & 24                                & 4                                 & 12                               & 12                               & 24                                \\
\textbf{readout rows}                         & \multicolumn{4}{c}{80 with 2-level addressing}  & \multicolumn{4}{|c}{80 with 2-level addressing}    \\      
\hline
\end{tabular}
\vspace{1mm}
\caption{Summary of detector and readout counts for the eight wafer types of CMB-S4. All wafers are dichroic except the \textit{ULF} wafer, with the detectors split evenly between frequencies. Total active detectors include a small number of dark channels for calibration purposes, and is the total across all telescopes.
}
\label{tab:ubertable}
\end{table}

\subsection{Integrated Detector \& Readout Module at the 100\,mK Focal Plane}\label{sec:100mK}

\begin{figure}[h!]
\centering
\includegraphics[width=0.85\textwidth]{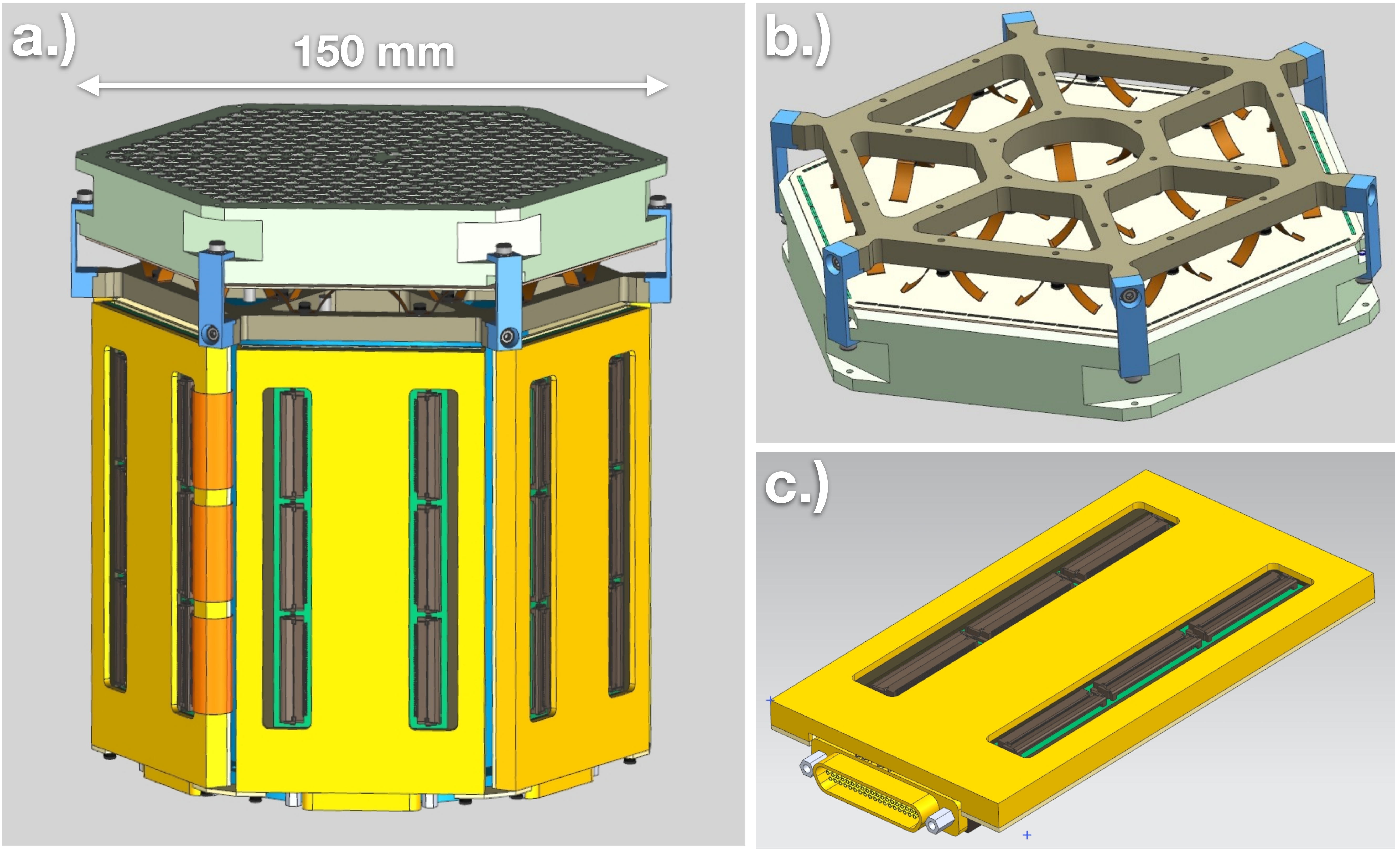}
\caption{3D model of integrated module of detectors, optical coupling, and readout electronics that will be installed in the telescope focal planes.
\emph{a.)} Fully assembled module.
The 100\,mK readout is mounted in magnetically shielded enclosures behind the footprint of the feedhorn arrays, and TES on the detector wafer are connected to readout by superconducting flexible circuits (not shown) wirebonded on both ends.
In the SATs, the hexagonal modules are tiled together, upto 12 in an optics tube, with a $\sim 1$\,mm gap between neighboring modules.
In the LATs, the modules are mounted one per optics tube.
The module shown is of the LAT MF design; other bands are similar, but use different feedhorn arrays and numbers of 100\,mK readout modules behind the optical coupling stack.
\emph{b.)} Assembled feedhorn and wafer stack
The detector stack wafers are aligned with a pin and slot system and clamped to the horn array with springs.
\emph{c.)} 100\,mK readout module.
The TES voltage bias and first-stage SQUID bias and feedback signals are routed to a micro-d connector, while row-select signals enter and exit the board on ZIF connectors that are exposed through the board housing.}
\label{fig:modulemontage}
\end{figure}

The detectors, readout electronics, and optical coupling components are mechanically integrated together in a module designed to operate at 100\,mK, which also provides mechanical and thermal interfaces to the cryostat and mixing chamber stage of the dilution refrigerator.
The design is driven by several key requirements: 1.) integrated modules must tile in the SAT focal planes with minimal deadspace, including $\lesssim 1$\,mm gaps between feedhorn arrays and 2\,mm wall thickness on the perimeter of the feedhorn; 2.) 100\,mK readout components must be arranged in identical, individually shielded, connectorized modules, each of which connects TES signals to the detector wafer via superconducting flexible circuits; 3.) the integrated module must have sufficiently high thermal conductivity both internally and to the cryostat thermal bus such that detectors can be stably operated at 100\,mK. 
Figure~\ref{fig:modulemontage} shows the key elements of the integrated module design.

The primary structural element of the integrated module is a copper ``spider plate'', shown in Figure~\ref{fig:modulemontage}, which clamps the optical coupling wafers onto the feedhorn array and mechanically supports the feedhorn array with tabs in each corner of the array.
Cutouts in the corner of the feedhorn arrays permit the use of low-profile tabs that enable modules to be tiled in the SAT focal planes with a  separation of $\sim 1$\,mm.
The secondary purpose of the spider plate is to provide a mounting point for the 100~mK readout modules.
The requirement that these modules fit behind the footprint of the feedhorn array imposes strict requirements on their width and motivates the use of 80-row multiplexed readout with only 4 readout columns per readout module.   There would be one readout module mounted per hexagonal side of the LAT MF, HF, and SAT HF integrated modules, which have the highest detector densities.
Since little space is available between the integrated modules for mounting and support structures in the SAT cryostat focal plane, modules are mounted to the focal plane structure by a tube that connects with the spider plate in the center space between the 100mK readout modules.
Since the LAT cryostat possesses individual circular optics tubes, there is more clearance than in the SAT around the perimeter of the feedhorn array for mounting features, and the integrated modules can be mounted to the cryostat structure either at the feedhorn or from behind as in the SAT.
Detailed designs of the module mounting are currently being developed in parallel with the LAT and SAT cryostat designs.

\subsubsection{150-mm Hexagonal Arrays of TES bolometers}


A detector wafer consists of a close-packed array of OMT pixels and TES bolometers. CMB detector wafers are designed for ease of mass fabrication via deposition, etch, and patterning of metals and dielectrics on a monolithic silicon substrate, and with special attention to high yield and uniformity across the array. CMB-S4's large quantity of detector wafers also necessitates control of processes over hundreds of wafers. A CMB-S4 detector wafer is a chamfered hexagon cut out of a $\sim500\,\upmu$m-thick 150-mm-diameter silicon wafer with 12 to 469 OMT pixels, depending on the wafer type, connected to 4 TES bolometers each. TES bias connections from each pixel are routed to the edges of the hexagonal wafer and presented as wire bond pads for connections to the 100\,mK readout modules via superconducting traces in a flexible circuit. 

\begin{figure}[h!]
\centering
\includegraphics[width=0.3\textwidth]{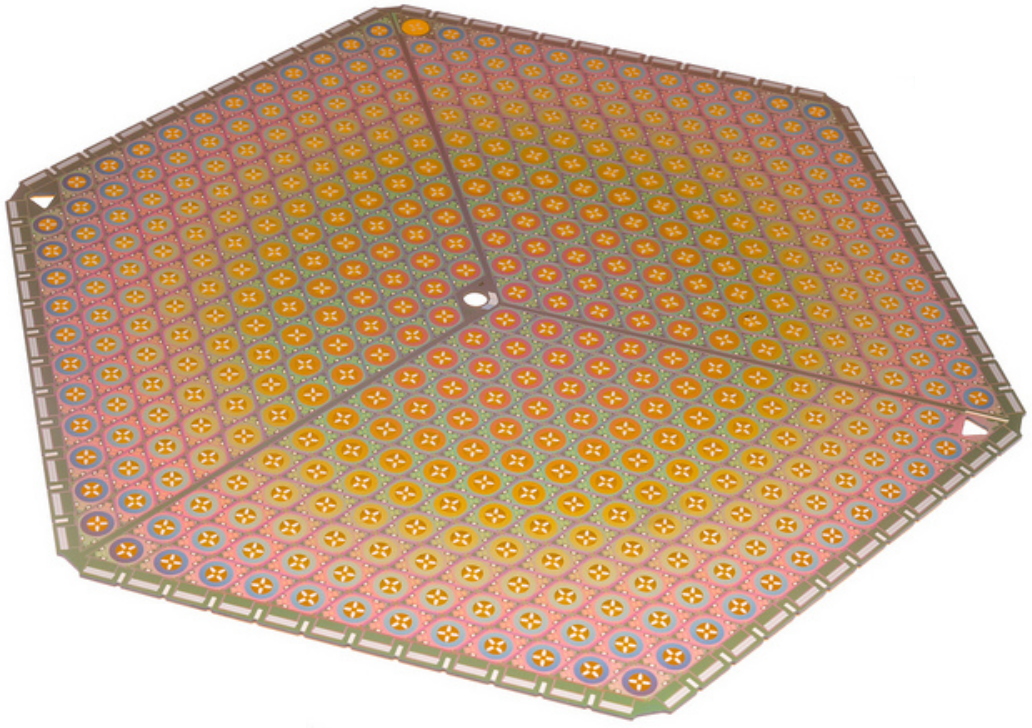}
\includegraphics[width=0.3\textwidth]{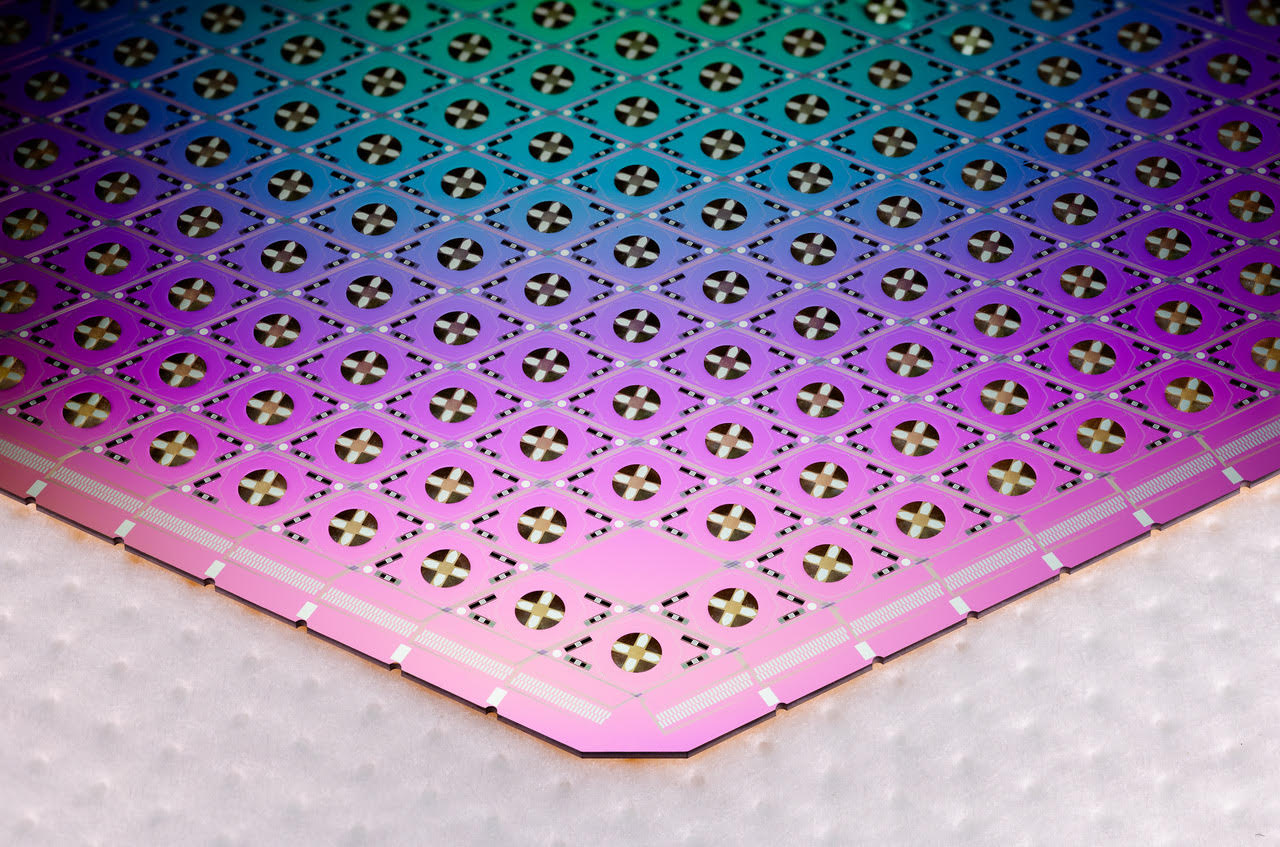}
\includegraphics[width=0.3\textwidth]{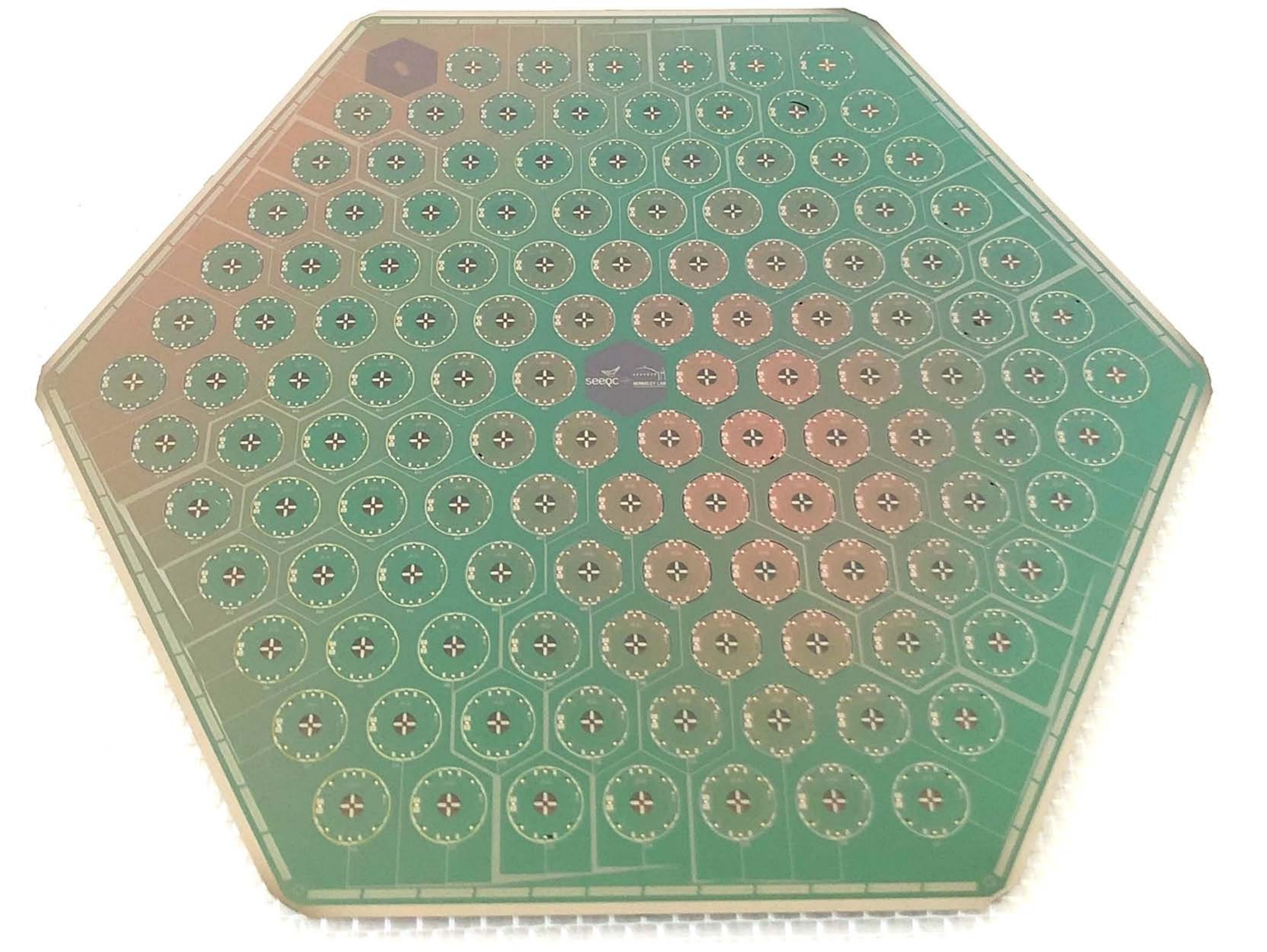}
\caption{{\it Left to Right:} 150-mm hexagonal detector wafer from NIST Boulder deployed in AdvACT and basis of CMB-S4 prototype, prototype from Argonne National Laboratory, and prototype from SeeQC/Lawrence Berkeley National Laboratory.}
\label{fig:teswafers}
\end{figure}

The detector wafer architecture used by CMB-S4 consisting of arrays of feedhorn-coupled OMTs coupled to TESs with $T_c \sim 160$\,mK has an extensive heritage, having been demonstrated by multiple experiments with high yield and uniformity.
A typical pixel operability of $>95\%$ has been achieved per wafer, with similarly high wafer-to-wafer yields \cite{DuffYield}.
In previous realizations of this process, deposition, etch, and pattern uniformity require much better than $\pm 5\%$ (shown to be improved to $<\pm 1\%$ for some processes) to achieve acceptable yield and performance; combined with high-quality pixel component designs, this results in the ability to achieve high efficiencies with very small spreads (as low as $\pm5\%$ across an array).
The detector fabrication processes for these devices, historically developed at NIST Boulder, are currently in the process of being ported to multiple microfabrication foundries in order to meet the immense fabrication throughput requirements of CMB-S4.
Due to the heterogeneous equipment available at the participating foundries, each site defines its own processes and choice of materials that meet the basic functional specifications of the CMB-S4 detectors. Some early prototype wafers are pictured in Figure~\ref{fig:teswafers}.

To integrate with the rest of the module, the detector wafers must conform to several mechanical interfaces.
First, the dimensions of the OMTs themselves are optimized together with the optical coupling in order to maximize efficiency subject to constraints on beam size, ellipticity, and polarization purity.
The layout of detectors within each wafer is optimized in order to meet both the NET requirement of the entire focal plane, together with the requirement on the maximum allowable spillover onto the cryogenic stop inside the telescope.
A mixture of two layouts, one with fully hex-close-packed (HCP) pixels and another with three, offset rhombus-shaped sections of HCP pixels are used for different frequencies of detectors.
The two layouts offer more flexibility in the number of detectors per wafer, while also providing flexibility to the fabrication sites in the style of wiring that is used (the rhombus layout enables creating wiring with stepper lithography).
Finally, the layout of bond pads on the perimeter of the wafer is standardized to be compatible with stepper-based wiring, using groupings of 25 pairs of 70-$\upmu$m wide, double-row bond pads.
The different wafer designs populate a subset of the groupings and a subset of pads within each grouping in order to maintain compatibility with the 100\,mK readout modules.

\subsubsection{Feedhorn to TES wafer stackup}
\begin{figure}[h!]
\centering
\includegraphics[width=0.99\textwidth]{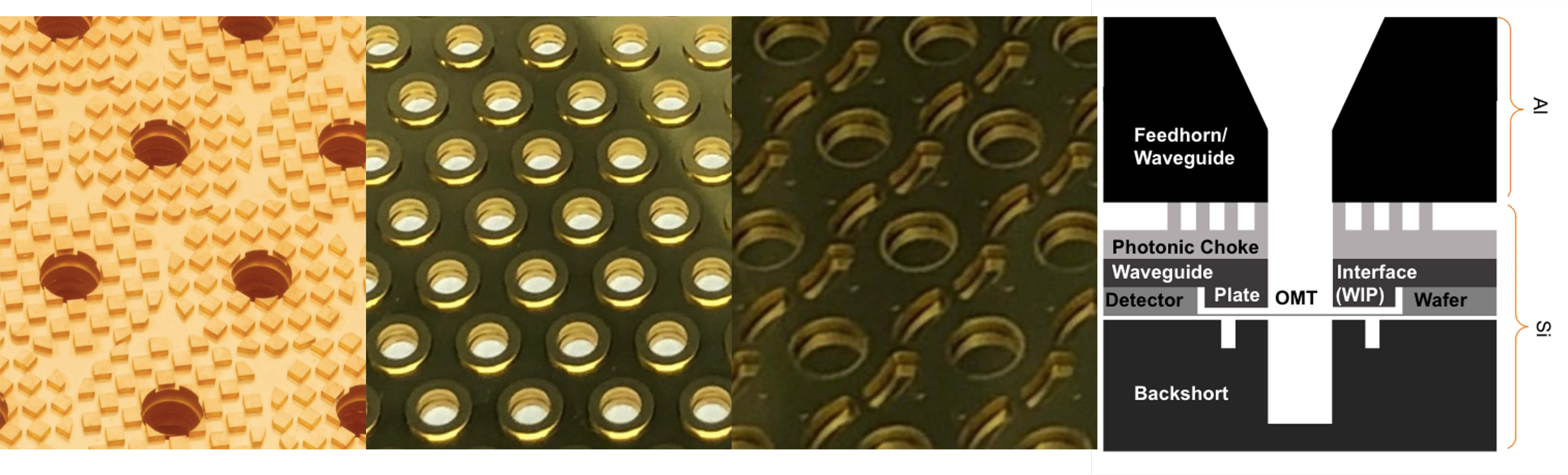}
\caption{{\it Left to Right:} Close up of the photonic choke wafer, waveguide interface plate wafer, backshort wafer, and a schematic cross section through the stack up of the optical coupling components and detector array.}
\label{fig:OC_stack}
\end{figure}

The Si detector stack is composed of a photonic choke wafer, a waveguide interface plate (WIP) wafer, the detector wafer, and a backshort wafer as shown in Figure~\ref{fig:OC_stack}. The photonic choke wafer has a pattern of square pillars that is optimized for each observing band to minimize leakage at the interface between the Al feedhorn array and the Si detector stack~\cite{photonic_choke}. The WIP has a ring-shaped boss feature on the backside of the detector wafer. The inner radius of the boss feature matches the waveguide radius of the detector stack and keeps the waveguide gap between the WIP and the OMT $<15\,\upmu$m. The outer radius of the ring is tuned to minimize leakage from the gap between the OMT and waveguide. The roughly quarter-wave backshort is tuned to optimize efficiency across the two bands. The backshort includes $10\,\upmu$m tall posts that offset the backshort wafer from the detector wafer wiring. The backshort also includes moats filled with absorptive material positioned behind the optical TES bolometers to reduce high frequency out-of-band leakage.

The optical coupling wafers are fabricated using silicon-on-insulator (SOI) wafers to tune the depths of the features. The features are etched into the wafers using DRIE, and then the wafers are seed-coated with 200\,nm of Ti and $1\,\upmu$m of Cu via sputtering to ensure even sidewall coverage. Next, the WIP and choke wafers are glued together into one piece, and the WIP+choke piece and backshort wafer are Au-coated with $3\,\upmu$m Cu followed by $3\,\upmu$m Au. After Au-coating, the coupling wafers are assembled together with the detector array. The pieces are placed in a simple gluing jig with two pins for alignment, clamped together, and glued in 2--3 glue channels on each side. The Si detector stack is coupled to the Al feedhorn array with a pin and slot system to account for the differential thermal contraction between the two materials when cooled. In the final assembly, the pins are press fit into the horn array, and the detector wafer stack is clamped to the feedhorn array with springs as shown in Figure~\ref{fig:modulemontage}. The feedhorn positions are oversized such that the two pieces are aligned when cold. The alignment tolerance of the mid-frequency waveguides is $20\,\upmu$m.

\subsubsection{Readout components at the 100 mK focal plane}

\begin{figure}[h!]
\centering

\tabskip=0pt
\valign{#\cr
  \hbox{%
    \begin{subfigure}[b]{.55\textwidth}
    \centering
    \includegraphics[width=\textwidth]{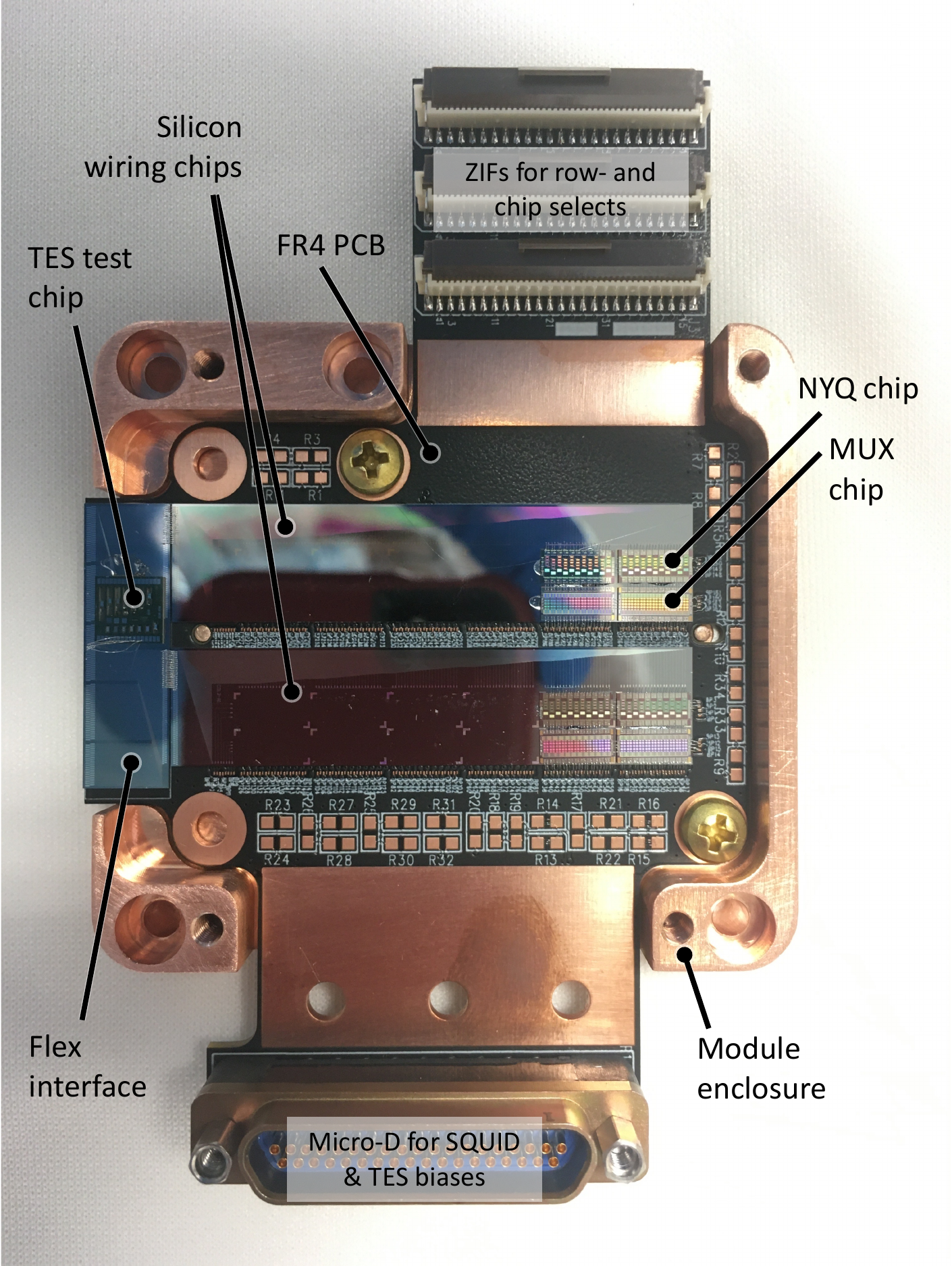}
    \end{subfigure}%
  }\cr
  \noalign{\hfill}
  \hbox{%
    \begin{subfigure}{.42\textwidth}
    \centering
    \includegraphics[width=\textwidth]{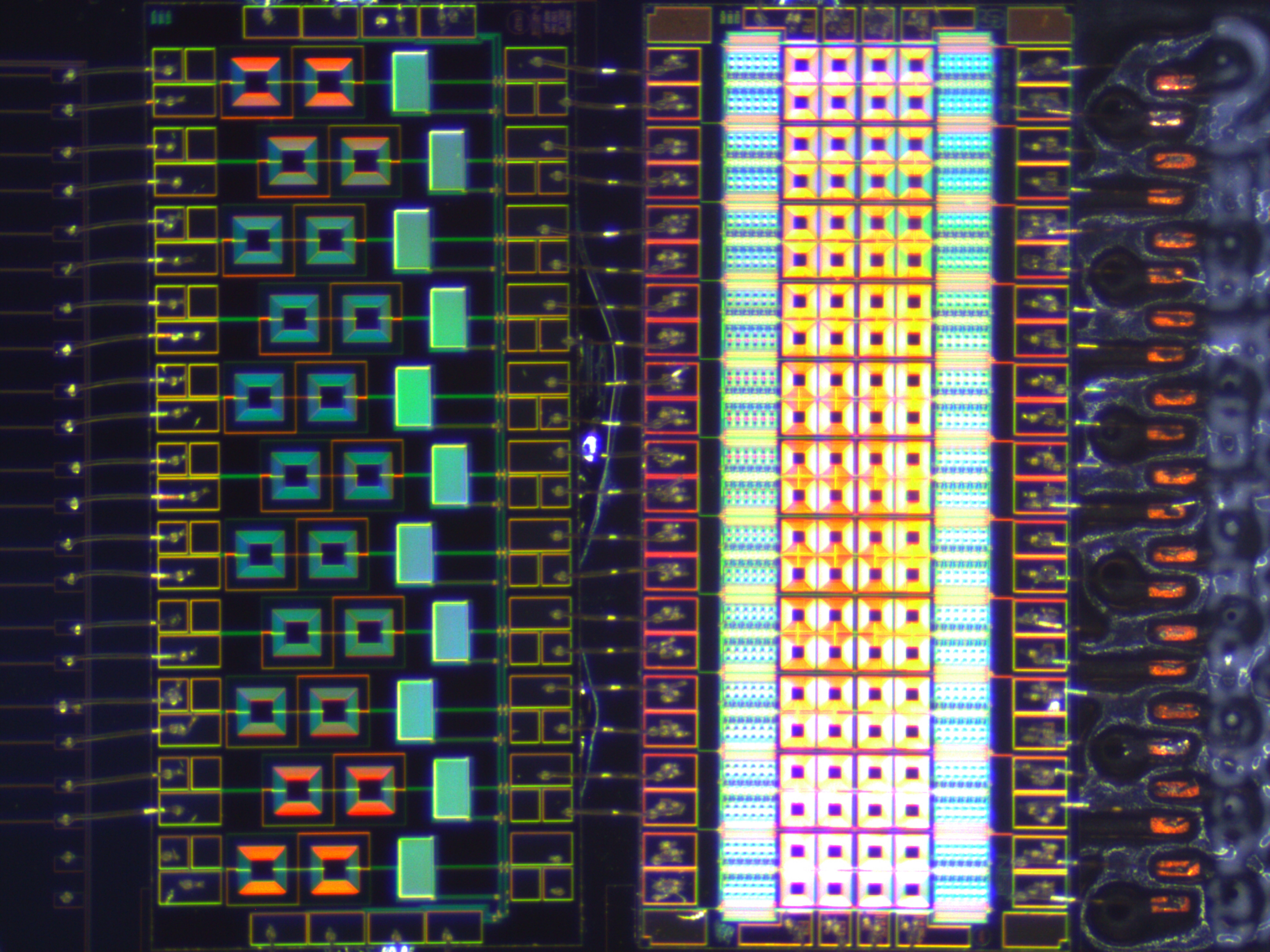}
    \end{subfigure}%
  }\vfill
  \hbox{%
    \begin{subfigure}{.42\textwidth}
    \centering
    \includegraphics[width=\textwidth]{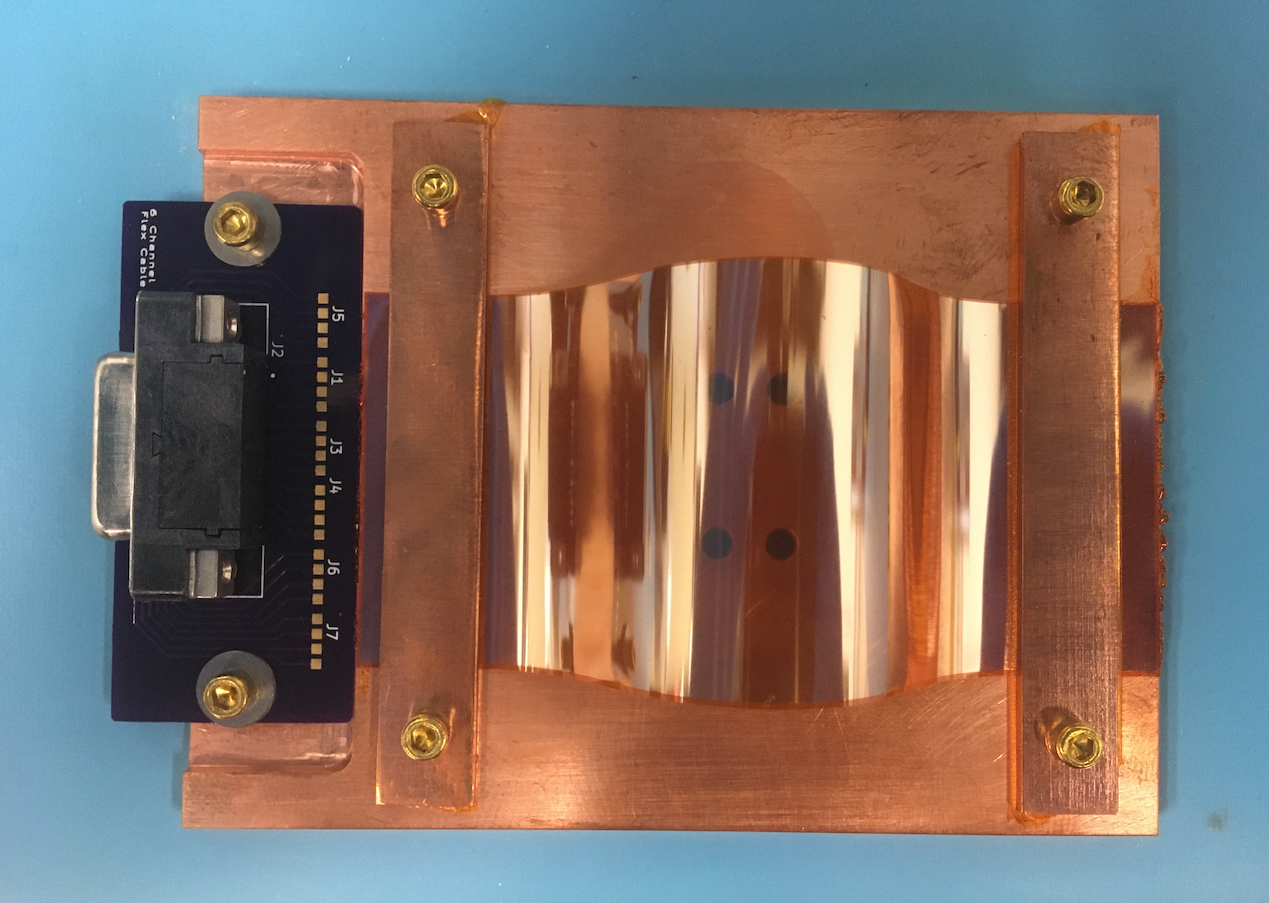}
    \end{subfigure}%
  }\cr
}
\vspace*{3mm}
\caption{
{\it Left:} Assembled prototype 2-column 100\,mK CMB-S4 readout module that houses the first stage SQUID amplifier and multiplexer (MUX) chips and the TES biasing (NYQ) chips. 
The MUX and NYQ chips are affixed to the surface of large silicon chips with superconducting wiring which will route signals from the TESs into the inputs of the NYQ chips.  The wiring chips will connect to the TESs (on the left) through a flexible superconducting circuit (not shown).
The silicon wiring chips are glued to the top surface of an FR4 PCB with ZIF connectors (top) for row- and chip-select inputs and a micro-D connector (bottom) for SQUID and TES biases.
The PCB housing (copper in this photo) is 55~mm x 60~mm.
{\it Top Right:} A pair of MUX (right) and NYQ (left) chips.  Readout circuits are completed using pairs of aluminum wirebonds and run horizontally across the chips.  Both chips are 3~mm x 6.6~mm.
%
{\it Bottom Right:} Flexible superconducting circuit prototype from SLAC, containing aluminum traces with 90\,$\upmu$m trace pitch on a thin polymide substrate. 
}
\label{fig:ro100mk}
\end{figure}

The first-stage SQUID amplifiers, multiplexing, TES biasing and signal filtering components are contained in 100\,mK readout modules co-located with the detector wafers at the focal plane. They are electrically connected to the wafer via a short flexible circuit with superconducting traces to limit parasitic impedances between the TES detector and the first stage amplifier. 
The first-stage SQUID amplifier (SQ1) and flux-activated row-select switches are fabricated onto a multiplexer chip (MUX) serving $\sim10$ TES channels each.The shunt resistors for TES voltage-biasing and the series inductors that define the TES signal and noise bandwidth are fabricated onto a ``Nyquist'' filter chip (NYQ), also called a TES biasing chip. 
The MUX and NYQ chips for a readout column are seated on and bonded to a larger silicon wiring chip. This chip contains superconducting traces that connect to the Nyquist chips on one edge, and present bond pads for connection to the TES detectors on a perpendicular edge. Sets of chips for four columns are assembled into a readout module. The components and their assembly into a prototype 2-column 100\,mK readout module is shown in Figure \ref{fig:ro100mk}. The 100\,mK readout module also receives row-select (RS) and chip-select (CS) control signals for multiplexing. 
Up to six readout modules, corresponding to 24 columns, are connected to a single row address board which is also located at the 100\,mK focal plane. 
This simple PCB distributes the RS and CS signals to the 100\,mK readout modules along flexible circuits with copper traces.



\subsubsection{Electrical interconnects and mechanical assembly}\label{sec:flexandassembly}

The requirement that the 100\,mK readout be modular practically necessitates the use of superconducting flexible circuits between the TESs and the readout.
The per-channel MUX and NYQ chip superconducting circuitry footprint is insufficiently compact to fit readout for entire SAT and LAT HF detector wafers on a single layer of Si directly behind the TES wafer.
Stacking multiple layers of Si readout components behind the TES wafer to accommodate the readout components would significantly increase the effort required for rework and replacement of components in the lower layers. Thus, a single-layer design was chosen for the 100\,mK readout modules, and the area required for Si components makes a superconducting flexible circuit an attractive option for routing the TES connections from the detector wafer to the readout.
Superconducting flexible circuits have been used in small quantities for this interconnect in multiple generations of CMB experiments using TDM, including ACTpol~\cite{ACTPol_Instrument}, AdvACT~\cite{Pappas:2016}, and CLASS~\cite{Dahal:2020} using Al traces.
Nevertheless, these circuits for CMB-S4 present several challenges including the large quantity of cables ($\sim$4,500 including spares), 90\,$\upmu$m trace pitch, and work hardening property of Al which limits the number of cycles of flexing.
R\&D is currently in progress on two fabrication processes.
The first uses Al film evaporated on kapton and then patterned with a lift-off process similar to one developed at SLAC National Accelerator Laboratory (SLAC)~\cite{Tomada:2015dia}.
The second process, developed by HighTec\footnote{\url{https://hightec.ch/}}, uses Nb patterned on a polyimide substrate and has achieved 10\,$\upmu$m feature resolution~\cite{Broise:2018}.

The final steps of the integrated module assembly consist of wirebonding the superconducting flexible circuit between the TES wafer in the feedhorn/wafer stack and the 100\,mK readout modules, and then mounting the readout modules on the back of the spider plate.
Similar to the design of AdvACT and SO modules, the CMB-S4 modules use Au wirebonds connected from Pd pads on the detector wafer to the Au-plated feedhorn arrays, to provide heatsinking.
After these are added, a Si DC wiring wafer is added on top of the backshort, and Al wirebonds carrying TES signals are added from the detector wafer to the DC wiring wafer.
This wafer serves as an adapter between the bondpad layouts on the flex cable and the detector wafer, and it allows the flex cables to route radially outward on the edge of each hex, which significantly simplifies the assembly process.
Superconducting flex cables are then glued to the DC wiring wafer and wirebonds are added.
The 100\,mK readout modules are finally folded behind the wafer and bolted to the spider plate, completing the assembly.

\subsection{Supporting cryogenic readout electronics}

\begin{figure}[h!]
\centering
\includegraphics[width=0.22\textwidth]{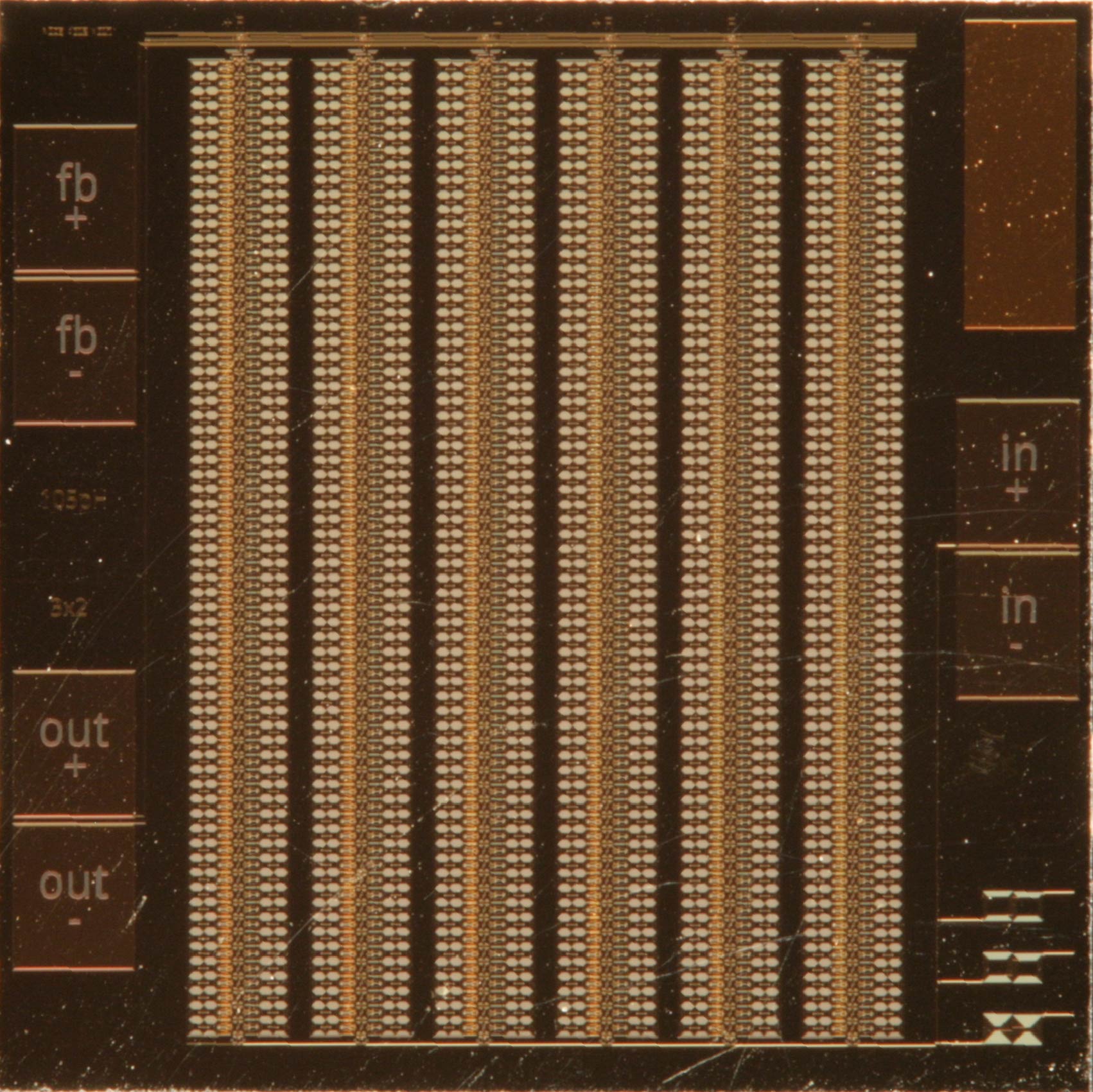}
\includegraphics[width=0.33\textwidth]{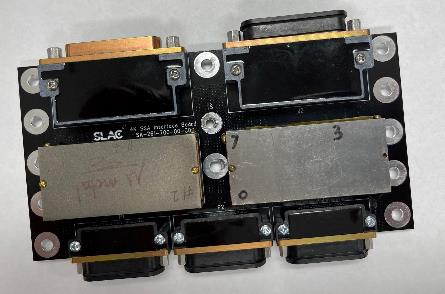}
\includegraphics[width=0.15\textwidth]{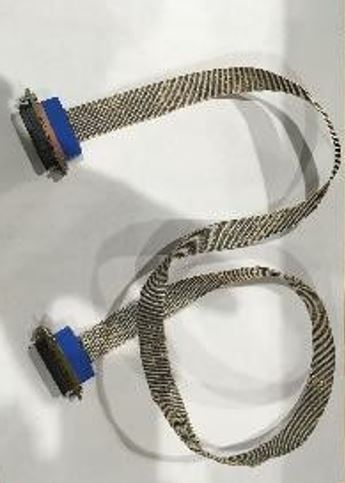}
\includegraphics[width=0.2\textwidth]{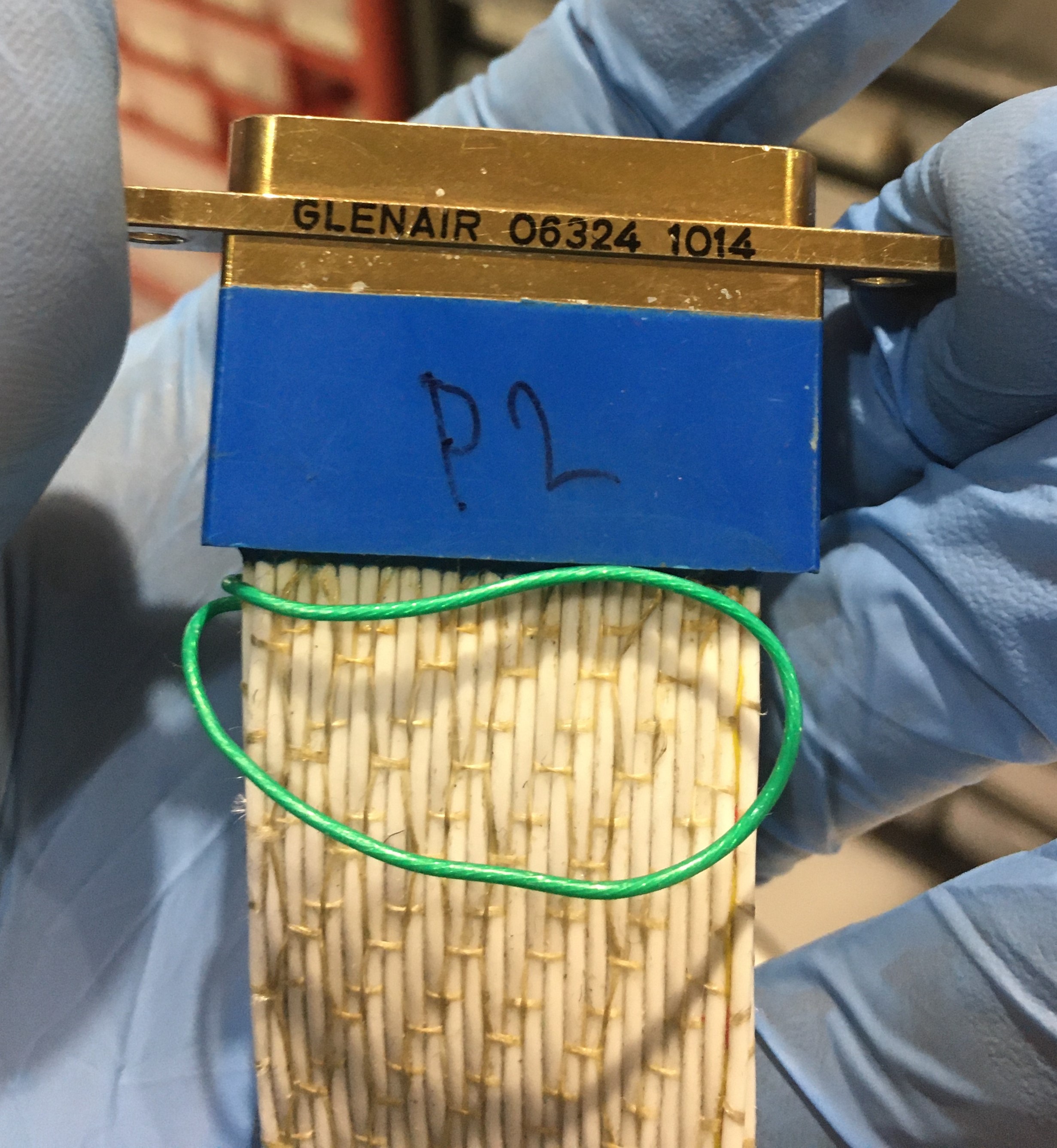}
\caption{{\it Left to Right:} A SQUID series array (SSA) with 6 banks of 64 SQUID elements, used for second-stage amplification of signals from the 100\,mK readout module; a prototype modular package with two sets of 8 SSAs under magnetic shielding; a low-thermal-conductance twisted-pair NbTi cable for connecting readout from the 100 mK stage to an intermediate temperature; a 100-wire shielded  twisted-pair manganin cable for connecting the cryogenic 8-column package to the room temperature electronics.}
\label{fig:4kelectronics}
\end{figure}

Signals from the integrated detector and readout module at 100\,mK described above continue to an additional SQUID amplification stage that must be located at a warmer temperature of the receiver, and eventually out to the room temperature readout electronics which control the multiplexing signals, provide detector and SQUID biases, and digitize detector signals. 
A schematic overview of these supporting electronics is shown in Figure~\ref{fig:drm_schematic}, example component and wiring counts are given in Table~\ref{tab:numerology}, and photos of prototypes are shown in Figure~\ref{fig:4kelectronics}.

	\begin{table}
	\centering
	\begin{tabular}{l|l|l|}
	& \textbf{SAT} & \textbf{LAT}                                                                                                          \\
		 &  \textit{HF (Single Wafer)}                     & \textit{Full Camera}                       \\
		\hline
		\textbf{Optical TES}                            & 1872                   & 137,904                \\ 
		\textbf{Total Active TES}                            & 1884                   & 138,414                \\ 
		\textbf{100 mK readout boards (4 columns each)} & 6                      & 470                    \\ 
		\textbf{Column cables (25-pin, 100 mK - 4 K)}   & 6                      & 470                    \\ 
		\textbf{Row address boards}                     & 1                      & 85                     \\ 
		\textbf{Row cables (51-pin, 100 mK - 300 K)}    & 1                      & 85                     \\ 
		\textbf{SSA Modules (8 SSAs each)}              & 3                      & 239                    \\ 
		\textbf{Column cables (100-pin, 4 K - 300 K)}   & 3                      & 239                    \\ 
		\textbf{Column boards (room temperature)}       & 3                      & 239                    \\ 
\textbf{Row boards (room temperature)}          & 1                      & 85                     \\   			\hline
	\end{tabular}
		
	\caption{Example TDM readout quantities are enumerated for an SAT HF detector wafer and an entire LAT receiver (85 wafers). Total active TES count includes dark detectors for calibration.}
				
	\end{table}\label{tab:numerology}


Low-thermal-conductance twisted-pair wiring is used to connect the cold electronics to warm electronics, with thermal intercepts at all possible temperature stages.  
The electrical design of this wiring is a straightforward application of wiring in demonstrated TDM systems, with some engineering challenges in managing thermal loads while staying within the overall impedance budget, and designing long cable runs to connect across the large focal plane. Design improvements to increase available bandwidth are described in Section \ref{ssec:sqampchainandmux}. Connecting the 100\,mK integrated module to second stage amplifier at warmer temperature, a 25-wire twisted pair NbTi cable per readout module of 4 columns carries SQ1 bias, SQ1 feedback, and detector bias.
The control signals for the row address module at 100\,mK are carried by a 51-wire NbTi cable which supports up to six readout modules (24 columns).

A SQUID series array (SSA) further amplifies signals for transmission to a room-temperature amplifier.
With multiplexing, only a single SQ1 first-stage SQUID feeds the SSA at any given time.  
To achieve the necessary amplification, the SSA must contain a large number of SQUID elements, which drives the design for this stage. 
The resulting power dissipation precludes the placement of this SSA on any of the sub-Kelvin cryogenic stages. 
The expected operating temperature is between 1\,and 4\,K, separate from the 100\,mK electronics and connected via low thermal conductance twisted pair wiring. 
These large arrays of SQUIDs are extremely sensitive to magnetic field variations and gradients, but their placement away from the focal plane allows for relatively compact, effective magnetic shielding as part of their packaging. 
Groups of 8 SSAs are packaged together into a module, which supports 8 columns of readout. A 100-wire manganin twisted-pair cable for this 8-column package carries the SSA bias and feedback between this temperature stage and room temperature, along with the SQ1 bias and feedback, and detector biases to be transmitted to 100\,mK.
The SSA is a mature design from NIST Boulder, which is a configurable array with 6 banks of 64 SQUID elements that can be connected in series or in parallel. 
The array layout and external shunting can be adjusted to modify electrical properties including input and output impedance, which is used in optimizing the gain and bandwidth of the system. 

\subsection{Room Temperature Electronics}

The SQUID multiplexer and amplifier chains described in Section~\ref{ssec:sqampchainandmux} require warm electronics for control and read out.  In particular, warm electronics provide SQUID biases and feedback for the two stages of SQUID amplification per readout column, row-select flux biases, and TES biases. 
The warm electronics also operate each TES in a closed flux-locked loop and stream digitally filtered and downsampled data from the receiver to data acquisition for storage and subsequent analysis.  
For CMB-S4, a new warm electronics readout system is being developed at SLAC. Previous CMB observatories have used the Multi-Channel Electronics (MCE) developed for the SCUBA-2 experiment as warm readout for TDM SQUID multiplexing, but several key components of the MCE system have reached obsolescence~\cite{battistelli_functional_2008}.  
The new system takes advantage of miniaturization of electronics since the design of the MCE, to significantly shrink the size of the warm electronics and enable CMB-S4's planned high-channel-count receivers.  
The new electronics are based on an extendable, compact, module-based architecture.  
In this new architecture, each warm readout system is a collection of two distinct types of modules with identical mechanical footprints that mount directly to connectors on the vacuum wall of the CMB-S4 receiver cryostats.  
All modules forming a single warm readout system are networked together using ethernet cables with one module serving as the controller for the other modules which interfaces externally with off-receiver data acquisition.

\begin{figure}[h!]
\centering
\includegraphics[width=0.49\textwidth]{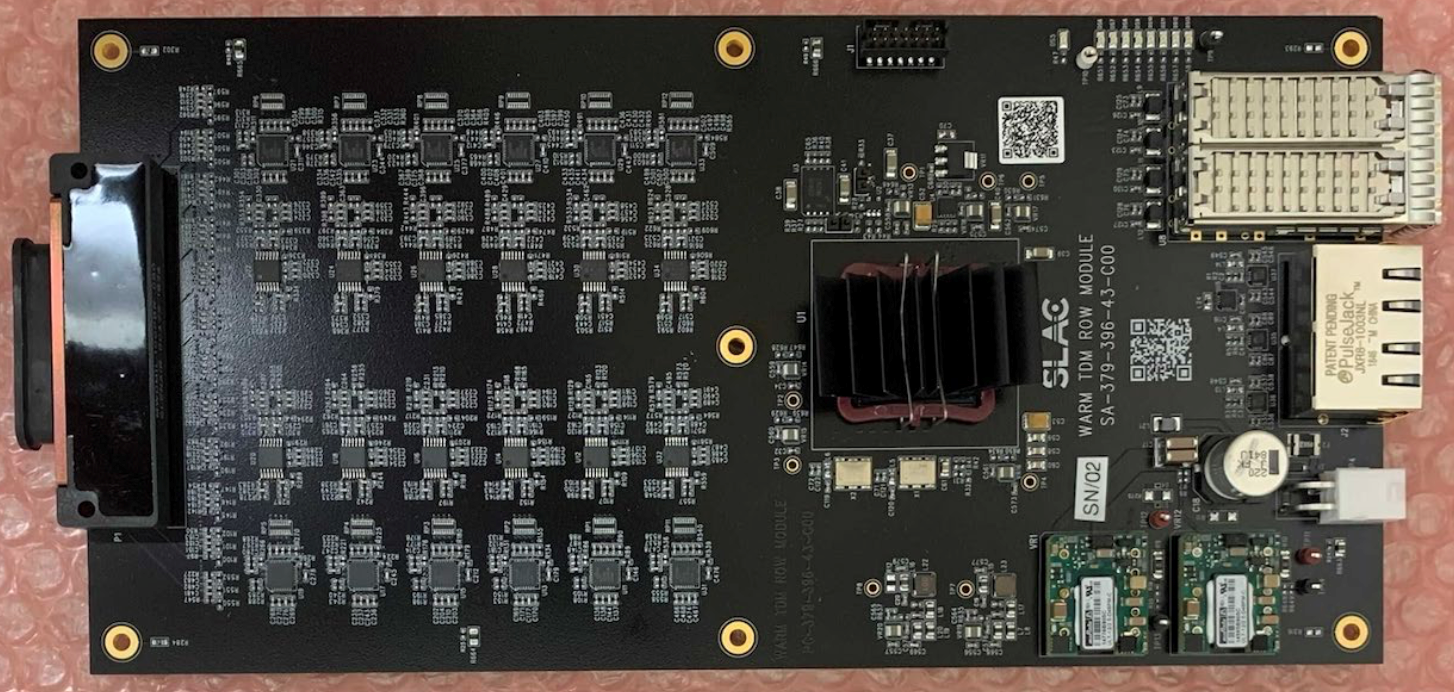}
\includegraphics[width=0.49\textwidth]{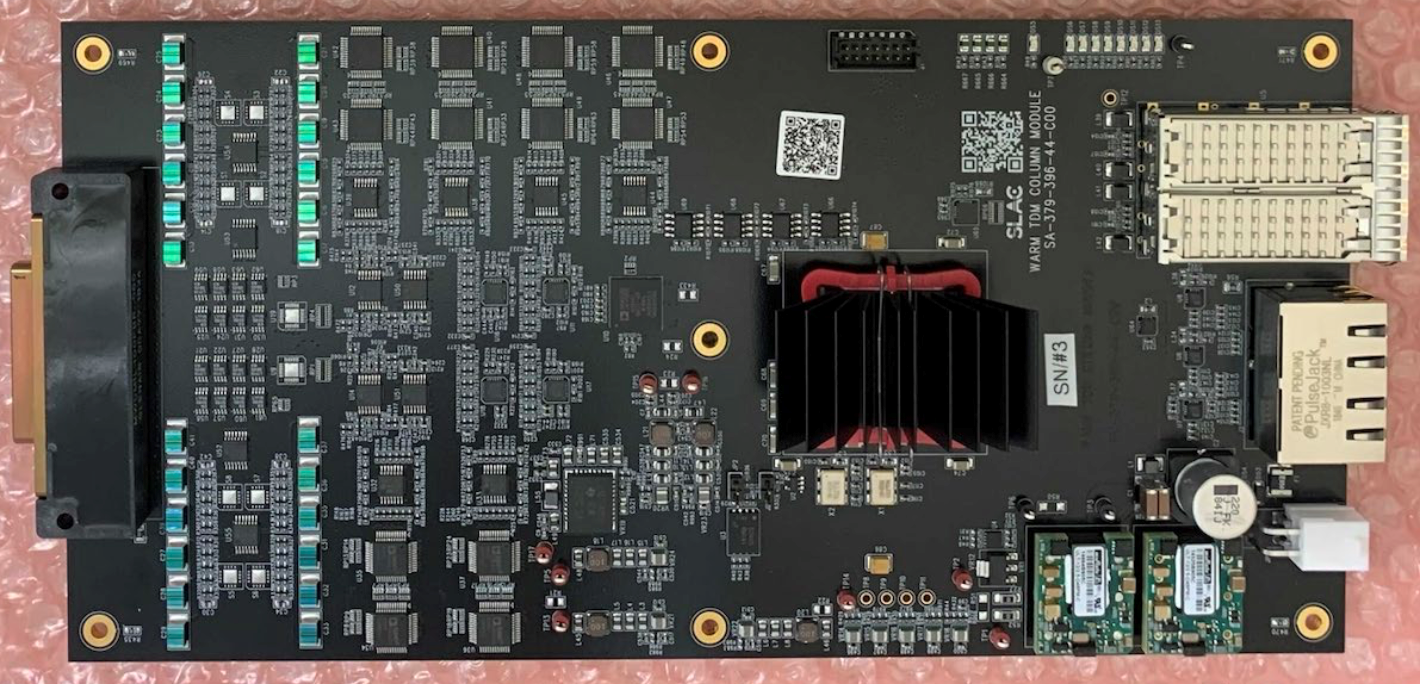}

\caption{Prototype warm readout modules developed for CMB-S4, shown here outside of their common housing.  Warm readout systems can be composed by chaining the two different types of modules, which have identical mechanical and electrical interfaces using ethernet cables, with one of the boards designated as the controller for a networked group.  Each module is designed to mount directly to connectors (on left) on vacuum flanges on the CMB-S4 receivers in a close packed configuration (not shown).  {\it Left:} A "row" type warm readout prototype module which can provide up to 48 row-switching signals to the 100\,mK readout modules. {\it Right:}.  A "column" type warm readout prototype module which can provide the TES biases, SQ1 and SSA bias and feedback signals, and low-noise analog front-end amplification for up to eight columns of a TDM SQUID multiplexer.}
\label{fig:slacwarmelexmodules}
\end{figure}

Each module connects to the CMB-S4 receivers via a male 100-pin micro-D connector with the same pinout as the legacy MCE system, to maintain backwards compatibility and enable testing with the MCE while the new system is under development.
The two types of modules are ``row" modules which each provide up to 48 row switching flux biases and ``column" modules which each provide the SQUID biases, SQUID feedbacks, and low noise analog front-ends for up to eight columns of TDM readout as well as up to eight TES biases.
Both row and column modules are 127~mm wide by 254~mm long, and have identical electrical back-end interfaces. 
In addition to being mechanically compact, the modules are designed to be conduction cooled enabling higher-density packing than comparable air-cooled systems, and eliminating the risk of microphonic and electrical pickup from fans.
On the opposite end of the modules from the 100-pin micro-D cryostat connector, both types of modules have a single 48V DC power input from which all other module voltages are derived using in-module regulators, two RJ-45 in/out connectors for networking groups of modules, and a dual Small Form Factor Pluggable (SFP) cage which supports both a 1~Gbps ethernet interface for testing and development and a timing input.
The row and column modules each have an FPGA controller which commands the analog-to-digital converters (ADC) and digital-to-analog converters (DAC) and handles digital processing tasks like operating each TES in its own closed flux-locked loop as well as data filtering, downsampling, and streaming.  Single-ended SQUID biases and feedback, row-select flux biases, and bipolar TES biases are all provided by DACs with in-module filtering to condition and bandwidth limit signals before they are injected into the cryostat.  Each of the eight low-noise analog front ends in the column modules consist of an amplifier chain with a low-voltage-noise first-stage preamplifier which feeds one channel of an integrated eight channel ADC.
The new electronics incorporate improvements informed by feedback from users of the legacy MCE system including fully integrated clock, timing, and communications, a single DC power input, a higher clock rate potentially enabling much higher multiplexing factors and data rates, and a modernized communication interface.

First prototypes of the row and column modules have been designed and assembled, shown in Figure~\ref{fig:slacwarmelexmodules}.  While fully functional, testing on SQUID multiplexers has indicated the need for a revision of the modules to address a few performance issues including higher than expected noise pickup from the in-module switching regulators used to step down the common 48V input voltage.
To help mitigate this and to improve performance generally, design changes are planned for this second revision including an improved filtering and grounding design for the switching regulators, fully differential SQUID biases and feedbacks, and a new lower noise, fully differential front-end design.

\section{Prototyping and Design Validation}
\label{sec:prototyping}

CMB-S4's development and design validation plan will mature the detector and readout system from a conceptual design to a prototype, and then to a preliminary design for pre-production before advancing to a final design for full production. This plan will advance both the integrated module sub-components, and the eight different integrated module types in a phased approach. The immediate goal during the next year of development is to demonstrate performance of the integrated detector and readout system with noise and optical coupling that meets the instrument requirements for a subset of module types, using prototype CMB-S4 hardware for a majority of the module components. We have developed readout and detector testing capability in cryogenic test stands, which will provide measurements of the module sub-components and integrated module performance (e.g., the TES properties on the detector wafer). Using one of these test stands, we have already conducted an end-to-end validation of the electrical design at the level of a few TES bolometers of different saturation powers connected to a TDM SQUID multiplexer. Further prototype development and measurements will be used to feedback and iterate on the design, and also validate that the sub-component requirements are still met in the integrated module. In this section, we describe the design and development plan for the prototype sub-components, and the planned testing of the integrated module.   

%
\subsection{Cryogenic Test stands for prototyping}\label{sec:test_stands}
%

\begin{figure}[h!]
\centering
\includegraphics[width=0.95\textwidth]{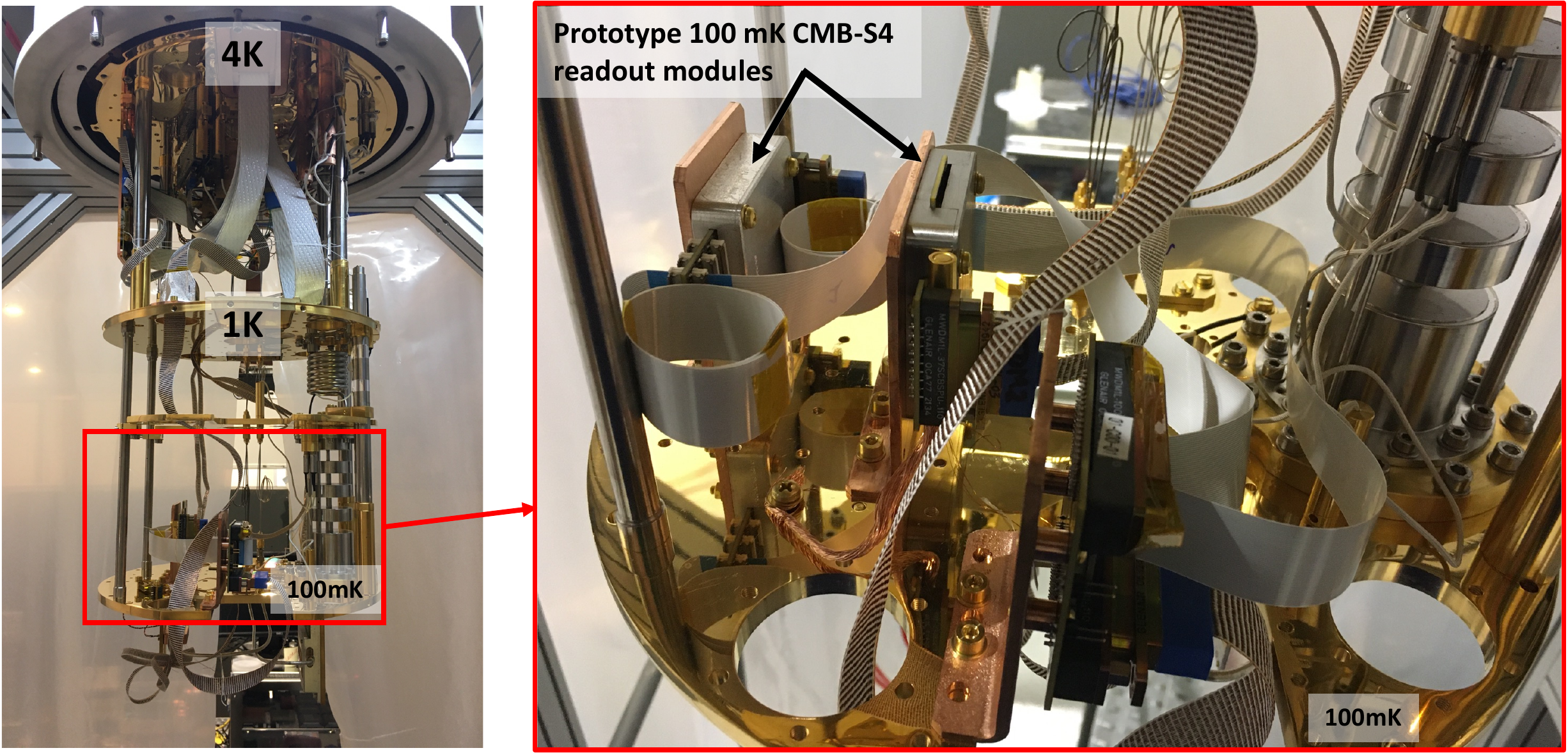}
\vspace{5mm}
\caption{{\it Left:} A CMB-S4 dilution refrigerator test stand at SLAC opened up to 4\,K showing the installed CMB-S4 TDM readout kit. {\it Right:}. Close up of prototype CMB-S4 100\,mK readout modules along with associated cabling and distribution boards installed for test on the 100\,mK mixing chamber plate.}
\label{fig:teststand}
\end{figure}

To support the prototyping of detector fabrication, readout electronics, and modules, the CMB-S4 project is commissioning three (eventually expanding to six) cryogenic prototyping test stands which are equipped with a standardized readout and test module design and  are each capable of testing a single 150-mm detector wafer.
These cryostats are either Bluefors LD-400 or Oxford Triton dilution refrigerators, made available to the project by members of the collaboration, which are equipped with legacy MCE room-temperature readout. We have designed and fabricated a complete, modular TDM readout kit which is drop-in compatible with these cryostats as shown in Figure~\ref{fig:teststand}.
The kit incorporates every component required from room temperature to 100\,mK including vacuum and thermal feedthroughs, cabling and fixturing, PCBs, and prototype 100\,mK readout modules, 4K SSAs and associated electronics.
Following an initial round of testing at SLAC, all systems were supplied with the kits.
This readout kit, in turn, is being used to develop a pre-prototype module (see Section \ref{sec:flatmodule}) at Fermi National Accelerator Laboratory (FNAL) that is being used to test prototype detector wafers from the TES wafer fabrication sites for CMB-S4.
In this early stage of project development, modules are wirebonded and assembled at FNAL and delivered to the other test stands in the project for testing.
A campaign of testing the same module repeatedly at multiple sites will provide a ``calibration'' ensuring that each site reports measurements comparable to the others.
The test stands will also provide feedback on prototype detector characterization equipment (see Section \ref{sec:testing}).
During pre-production, as readout and module designs are finalized, the project will commission eight high-throughput test cryostats, each capable of testing seven integrated detector and readout modules; these cryostats will supersede the prototyping test stands.

\subsection{Detector and Readout functional validation}

We have connected a few individual TES bolometers to our prototype TES biasing and readout implementation to successfully validate the electrical design of the detector and readout system. While a new advanced cryogenic TDM SQUID multiplexing architecture, described in Section~\ref{ssec:sqampchainandmux}, is being developed for CMB-S4, it is designed to be backwards compatible with the legacy TDM architecture used in currently active CMB observatories. This has enabled functional validation of many readout components using the legacy hardware.  Likewise the legacy multiplexer is being used to characterize and validate TES designs, from single devices to full wafers.  In particular, directly connecting individual TES bolometers to the multiplexer decouples readout design validation from that of the TES wafers and the integrated detector and readout module. 

As shown in Figure~\ref{fig:ro100mk}, single pre-screened, NIST first-stage SQUID 11-channel multiplexing chips, mask name ``mux15b", were integrated into prototype readout modules and directly connected to TES test devices through a prototype TES bias or ``shunt'' chip.  The TES biasing chip, connected between the multiplexing chip and TES devices with superconducting aluminum wirebonds, shunts the TESs with 450\,$\upmu\Omega$ bias resistors.  On the shunt chip, these bias resistors are connected in series on a common bias line which enables voltage biasing the TESs into transition.  The TES bolometers were fabricated at NIST Boulder, several  per test die, with saturation powers spanning the range of expected CMB-S4 TES device parameters (See Table~\ref{tab:ubertable}), from $\sim0.3-30$\,pW.  The TESs had normal resistances of $\sim10$~m$\Omega$ and transition temperatures of $\sim160$\,mK.

\begin{figure}[h!]
\centering
\includegraphics[width=0.8\textwidth]{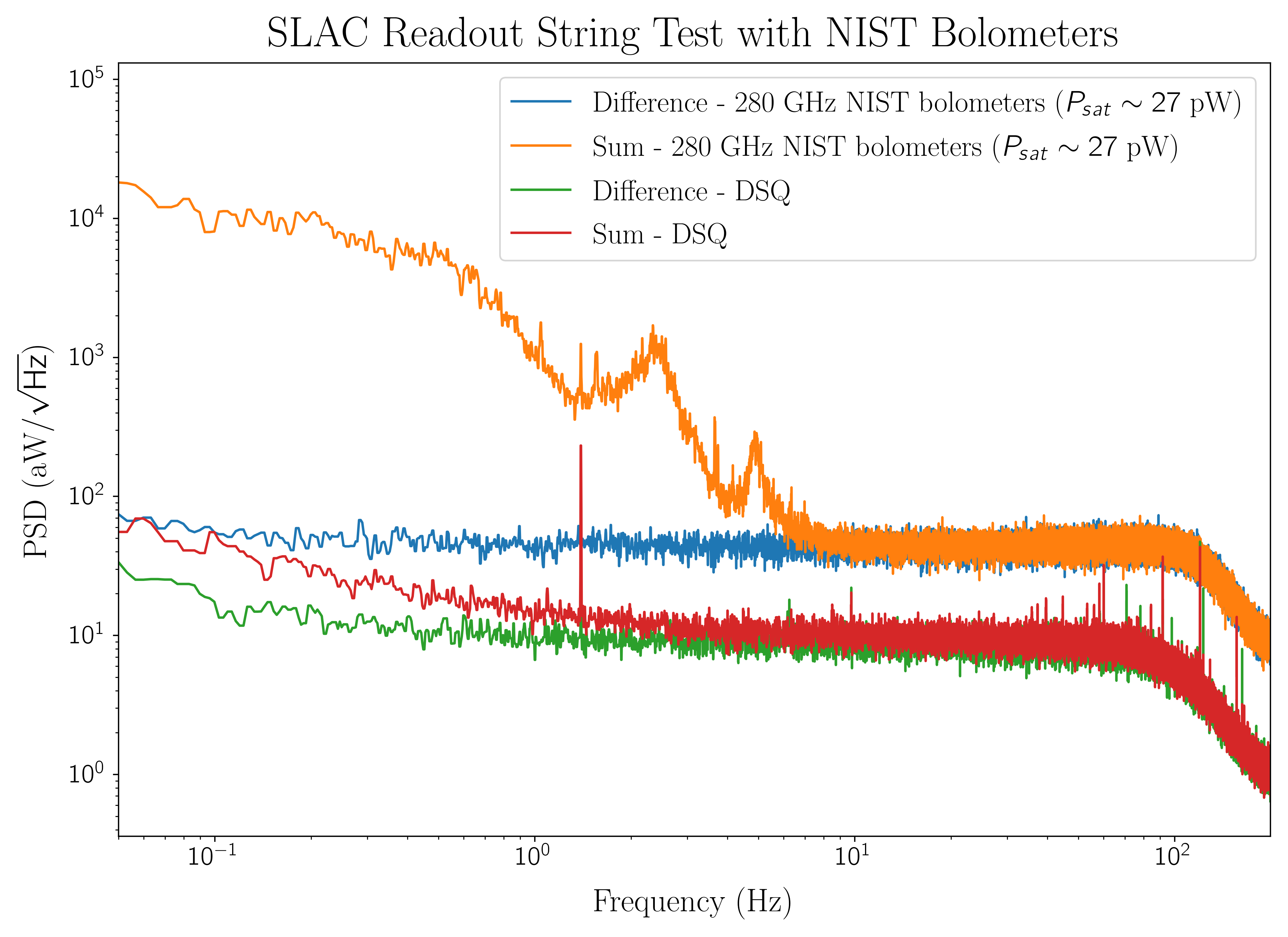}
\caption{Measured end-to-end noise-equivalent power of prototypical TES bolometers with the CMB-S4 modular TDM readout test kit installed in a dilution refrigerator test stand at SLAC.  Sum (orange) and differenced (blue) noise is shown for a pair of TESs from NIST Boulder voltage biased at 50\% of their normal resistance.  The TES pair are of identical design with saturation powers of $\sim27$~\,pW.  The measured noise has been converted to equivalent power noise in aW$/\sqrt{\mathrm{Hz}}$ by scaling the data from each TES by its responsivity estimated from an I-V sweep.  A small relative gain correction ($\lesssim 5$\%) has been applied to the noise spectrum of one of the TESs to match their quadratic means.  Noise power below $10$~\,Hz in the pair sum TES noise is differenced away, likely originating from fluctuations in the 100\,mK base temperature.  The white noise level in the pair difference TES noise is consistent with the intrinsic phonon noise expected in the devices.  Dark SQUID (DSQ) channels, whose first-stage SQUID inputs are left open, are used to measure readout noise alone, and the pair-differenced DSQ noise is nearly an order of magnitude lower than the pair differenced TES noise.  Here the DSQ noise has been scaled by the average of the estimated responsivities of the two TES devices.  The measured readout-only pair-differenced white noise level is in agreement with known performance.  Excess pair-differenced DSQ noise at low frequencies is likely related to the experimental setup and could be mitigated in future measurements.}
\label{fig:tesnoisewithrokit}
\end{figure}

Pairs of identical TES devices were connected to adjacent rows of the multiplexer to allow characterization of individual, pair sum, and pair difference noise.  Other adjacent inputs on the multiplexer were left unconnected, allowing for a measurement of the readout noise alone.
For these measurements, the MCE switched over 33 rows, with a row switching rate of 2\,$\upmu$s, even though only the first 11 rows were instrumented with first stage SQUIDs.
Figure~\ref{fig:tesnoisewithrokit} shows a comparison of the measured readout noise to noise measured on a pair of TES devices with a saturation power of $\sim27$~pW.
White noise performance of the readout is in agreement with known performance.  Excess noise at low frequencies is under investigation but thought to be due to a combination of the lack of temperature regulation in the laboratory space where this testing was conducted and known performance issues with the older revision legacy MCE boards used for these measurements.
%



%
\subsection{Flat Module Development}\label{sec:flatmodule}
%

\begin{figure}[h!]
\centering
\includegraphics[width=0.95\textwidth]{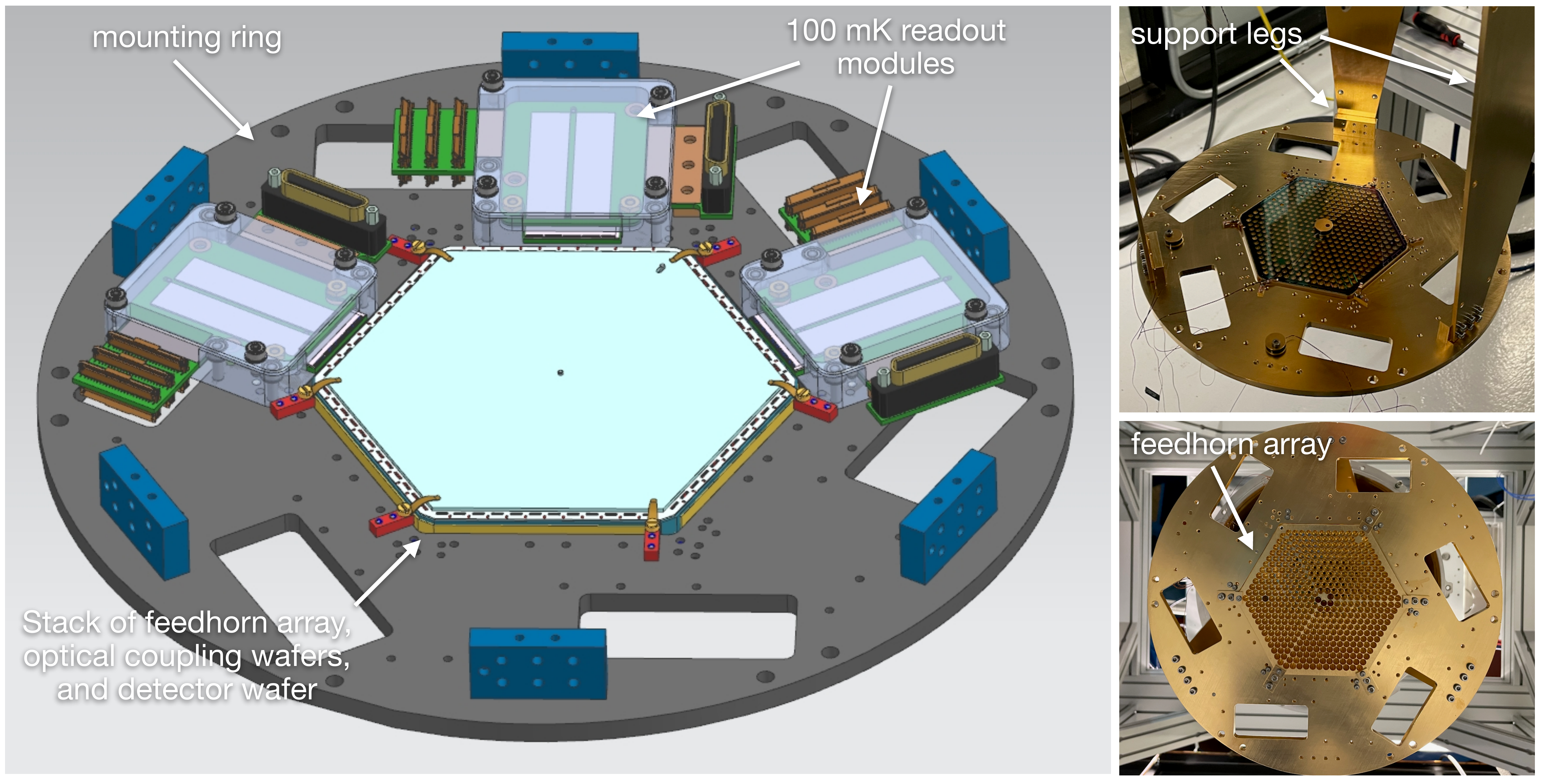}
\caption{Flat module used for prototyping of detectors, readout, and optical coupling components. \emph{Left:} 3D model of assembled module, showing the stack of feedhorn array, optical coupling wafers, and the detector wafer itself, which are held in place by spring clips. The 2-column 100~mK readout modules are mounted radially around the detectors and bolted onto a mounting ring that supports the assembly. \emph{Top right:} The flat module, installed in a CMB-S4 dilution refrigerator test stand at FNAL, suspended from the mixing chamber by adjustable legs. These legs allow the feedhorn array to be positioned in front of an cold load (see \ref{sec:testing}) or optical filter stack and vacuum window. \emph{Bottom right:} View of the installed flat module, looking at the feedhorn array.}
\label{fig:flatmodule}
\end{figure}

The next stage of development integrates the readout kits with prototype TES wafers in a pre-prototype ``flat module''. Due to the ongoing development of the superconducting flexible circuits described in Section~\ref{sec:flexandassembly}, integrated ``string'' tests of detectors and readout electronics use a scheme in which 100\,mK readout modules are arrayed radially on each side of the detector wafer, enabling pads on the detector wafer and the readout to be directly wirebonded without the use of flex cables, as shown in Figure~\ref{fig:flatmodule}.
This module fulfills the programmatic goal of performing full-system tests of CMB-S4 detectors and readout as early as possible in the project development cycle.
Using a geometry similar to the module of AdvACT~\cite{Ward:2016}, this design provides a platform for end-to-end testing of the entire system, both dark and optically, without the superconducting flexible circuit and with more relaxed space requirements than the production module.
The prototype 100\,mK readout modules, shown in Figure~\ref{fig:ro100mk}, use 2 readout columns per module, with each module bonded to a subset of the detectors on each side of the wafer and the option of installing one readout module on each side of the detector wafer.
Detector wafers may be tested with or without any optical coupling wafers, and in the latter configuration the detector wafer is simply secured to the feedhorn with brass spring clips.

\subsection{Dark and Optical characterization}\label{sec:testing}
Flat modules will undergo both dark and optical tests to compare with CMB-S4 targets and provide feedback to the TES wafer development program of CMB-S4. These tests will characterize the TES properties across the entire wafer as well as the integrated performance of the module components. The tests will be initially performed in the cryogenic test stands described in Section \ref{sec:test_stands}.

\begin{figure}[h!]
\centering
\includegraphics[width=0.8\textwidth]{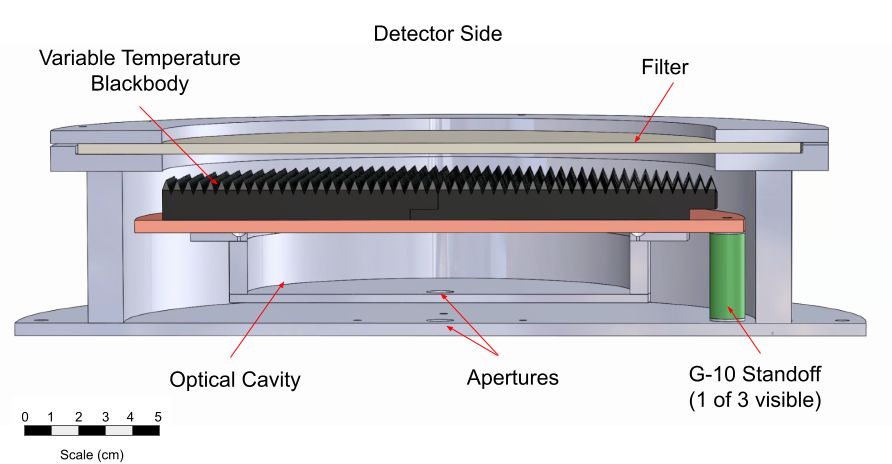}
\caption{A variable-temperature cold load with calibrated thermal sensors inside the cryostat will enable measurements of the optical efficiency and overall sensitivity of the integrated modules.  In this prototype design, a cavity with a small aperture behind the blackbody enables the detectors to view small optical signals to measure the optical time constants.}
\label{fig:coldload}
\end{figure}

The dark tests will characterize integrated module properties such as channel operability, TES properties, detector stability, and time constants.  Variable temperature blackbody sources will be used to estimate the module optical efficiency from the detector response.  The dark tests will also estimate the overall sensitivity of the detectors, or noise-equivalent power (NEP), from a combination in-transition current noise measurements and detector load curves at multiple blackbody source temperatures.   In the prototype calibrator design shown in Figure \ref{fig:coldload}, a flashing IR source coupled through a small aperture in the blackbody also enables measurements of the optical time constants of the science detectors.  These will be compared to the time constants inferred from the TES response to an electrical voltage bias step.

The optical characterization of the integrated modules will be performed by using equipment coupled to the cold  modules through a series of out-of-band radiation filters and a vacuum window. For instance, the detector frequency response will be measured using a Fourier transform spectrometer; low-pass and high-pass thick grill filters will be used to check for spurious out-of-band detector response.  Other optical properties will be spot-checked during development including beam shape, cross-polarization response, and polarization sensitivity angle.  RF and magnetic pickup of the integrated module will also be measured using swept RF and magnetic sources. 


\section{Future work towards production at scale}\label{sec:production}

After the prototyping phase, described in the previous section, the CMB-S4 project would have verified that the detector and readout system components, and their integrated system meet the Level 2 subsystem requirements, enabling the start of pre-production.  During pre-production, the project will demonstrate quality assurance, component fabrication, integration, testing, and quality control steps for a fraction of the required integrated modules and their components, but at the necessary rate and throughput for full production.  
Quality assurance and control are key in this phase, and a logging system
will be used to record key metrics that enable monitoring and control of the fabrication processes. During production the logging system data will be routinely reviewed to detect process variations and correct them before enough variation occurs to affect performance.

During production, the full set of approximately 500 science-grade modules will be delivered over a roughly 3-year period.  This will require producing and screening an estimated 700 TES wafers and 150 SQUID wafers, along with associated optical, readout, and module components.  This amounts to fabrication of complex superconducting circuits on over 10\,m$^2$ of silicon, as well as significant amounts of precision wiring, assembly and cryogenic testing.
To meet the required fabrication rate of approximately 20 TES wafers and 5 SQUID wafers per month will require a multi-site fabrication approach. Several micro- and nanofabrication foundries specializing in superconducting thin film fabrication will be utilized by the project to achieve this rate. We plan to consolidate 100\,mK integrated  module assembly and testing into 2 sites, in order to reduce the amount of duplication of expertise and infrastructure. Each testing site will house four high-throughput screening cryostats, capable of characterizing seven modules per cooldown at the nominal 100\,mK operating temperature. As done during prototyping and pre-production, each module will be tested twice, performing a series of dark and optical characterization measurements to verify that it is science-grade by meeting the instrument requirements.

\appendix    

\acknowledgments 
CMB-S4 is supported by the U.S. Department of Energy (DOE), Office of High Energy Physics (HEP) under Contract No. DE–AC02–05CH11231; by the National Energy Research Scientific Computing Center, a DOE Office of Science User Facility under the same contract; and by the Divisions of Physics and Astronomical Sciences and the Office of Polar Programs of the U.S. National Science Foundation under Mid-Scale Research Infrastructure award OPP-1935892. Work at Argonne National Lab including use of the Center for Nanoscale Materials, a U.S. Department of Energy Office of Science User Facility, was supported by DOE, Office of Basic Energy Sciences and HEP, under Contract No. DE-AC02-06CH11357. Work at the Fermi National Accelerator Laboratory, managed and operated by Fermi Research Alliance, LLC was supported by DOE HEP under Contract No. DE-AC02-07CH11359. Work at SLAC National Accelerator Laboratory was supported by DOE HEP under contract DE-AC02-76SF00515. D.R.B. was supported by DOE HEP under award number DE-SC0021435, and NSF's Office of Integrative Activities under award OIA-2033199. Considerable additional support is provided by the many CMB-S4 collaboration members and their institutions.  

\bibliography{report} 
\bibliographystyle{spiebib} 

\end{document}